\documentclass[reprint,superscriptaddress,amsmath,amssymb,aps,prc]{revtex4-2}

\usepackage{graphicx}
\usepackage{dcolumn}
\usepackage{bm}

\begin{document}

\title{Active-learning construction of hyperspherical-harmonics spaces for $A=3$}

\author{Elena Filandri}
\email{efilandri@ectstar.eu}
\affiliation{European Centre for Theoretical Studies in Nuclear Physics and
 Related Areas (ECT*), Fondazione Bruno Kessler, Strada delle Tabarelle 286,
 I-38123 Trento, Italy}

\author{Andrea Di Luca}
\affiliation{Data Science for Industry and Physics (DSIP), Fondazione Bruno
 Kessler, Via Sommarive 18, I-38123 Povo (Trento), Italy}

\author{Laura Elisa Marcucci}
\affiliation{Dipartimento di Fisica ``E.~Fermi'', Universit\`a di Pisa,
 Largo Bruno Pontecorvo 3, I-56127 Pisa, Italy}
\affiliation{INFN-Sezione di Pisa, Largo Bruno Pontecorvo 3,
 I-56127 Pisa, Italy}

\author{Michele Viviani}
\affiliation{INFN-Sezione di Pisa, Largo Bruno Pontecorvo 3,
 I-56127 Pisa, Italy}

\date{\today}

\begin{abstract}
The hyperspherical-harmonics (HH) expansion does not require uniform resolution: different components of the wave function converge at very different hyperangular and hyperradial scales. We use active learning to exploit this structure, allowing the calculation to distribute resolution instead of prescribing it. The HH space is decomposed into physically
identifiable classes whose hyperangular and hyperradial cutoffs evolve
independently; a Gaussian-process surrogate learns the marginal variational
gain of each admissible extension, and a cost-aware acquisition selects the
next one.  The surrogate never replaces the many-body solver: every accepted
extension is followed by an explicit solution in the enlarged space, so each
reported energy is variational in an explicitly constructed basis.  We test
the construction on $^3$H and $^3$He with the Argonne $v_{18}$ two-nucleon
interaction, without and with the Urbana IX three-nucleon interaction,
against uniform HH ladders extended to $K=60$ that reproduce
established benchmarks within $1$~keV.  Asked for $5$~keV, the adaptive runs
reproduce the energy of the uniform $K=60$ space within $4$~keV in every
case and certify it, leaving more than $60$\% of that space unbuilt.  The
reduction grows as the requested accuracy is relaxed, to about $88$\% at
$20$~keV and $95$\% at $100$~keV.  The selected spaces reproduce the known
class hierarchy of the trinucleon, and the one-body radii and magnetic
moments are converged at about the $2\times10^{-3}$ level (relative error) in
the energy-selected space.  A run started from the resolution reached by the mirror nucleus,
or by the same nucleus with the two-nucleon interaction alone, certifies
a smaller space with a third of the exact solves.
\end{abstract}

\maketitle

\section{\label{sec:introduction}Introduction}

Few-nucleon systems allow the Schr{\"o}dinger equation to be solved with very high numerical accuracy, making it possible to separate deficiencies of the underlying nuclear dynamics from uncertainties associated with the numerical solution. This makes light nuclei particularly useful for testing nuclear interactions and the operators entering calculations of bound-state properties, reactions, and transitions. Three- and four-nucleon systems have consequently played a central role in the development and validation of modern nuclear Hamiltonians, with benchmark calculations from independent methods agreeing at the few-keV level~\cite{Nogga2003,Kamada2001}.

The hyperspherical-harmonics (HH) method expands the nuclear wave function in
a basis with explicit orbital, spin, isospin and grand-angular quantum
numbers~\cite{Kievsky2008,Marcucci2020}.  It treats both local and nonlocal
interactions, has reached high accuracy for three- and four-nucleon
bound and scattering states, and has been extended to $A=6$~\cite{Gnech2020}.  Convergence is improved by raising the
hyperangular cutoff $K$ and enlarging the hyperradial expansion, at the price
of a variational space that grows rapidly with both the requested accuracy
and the number of particles.

That growth is largely spent on components that contribute little.  HH
convergence is strongly nonuniform: some parts of the wave function need a
large hyperangular resolution, others become irrelevant at small $K$.  The
property is long established~\cite{Viviani2005,Barnea2000}: convergence was
analyzed class by class in calculations of the $^4$He ground state, which
reached accurate results in a restricted rather than uniformly enlarged
basis~\cite{Viviani2005}, and the effective-interaction HH approach
accelerates convergence in a truncated space by constructing an effective
interaction for that space~\cite{Barnea2000}.  The pair-correlated HH
expansion reduces the space in yet another way, by building the short-range
correlations into the basis functions
themselves~\cite{Kievsky1993,Kievsky1994}.  Treating all components at
the same resolution is therefore unnecessary.

Identifying the relevant part of a rapidly growing Hilbert space is a general
problem shared by the ab initio methods that have extended the reach of
nuclear structure calculations, from the no-core shell model~\cite{Barrett2013}
to quantum Monte Carlo~\cite{Carlson2015}.  Importance-truncation schemes in
configuration-interaction calculations use perturbative estimates of the
relevance of individual basis states to build a reduced model space
iteratively~\cite{Roth2007,Roth2009}, as the selected
configuration-interaction methods of quantum chemistry
do~\cite{Huron1973,Holmes2016}.  Statistical and machine-learning
methods have meanwhile become part of the standard toolbox of the
field~\cite{Boehnlein2022}, from Bayesian analyses of few-nucleon
observables~\cite{Gnech2024} to reduced-basis emulators built by eigenvector
continuation~\cite{Frame2018,Melendez2021,Duguet2024}.
More recently, active learning has been combined with such emulators:
residual-based greedy algorithms select high-fidelity training points in the
parameter space of nuclear interactions~\cite{Maldonado2025}, and the same
strategy was applied to proton-deuteron scattering with HH calculations as
the high-fidelity solver~\cite{Gnech2025}.

In this work active learning plays a complementary role: for a fixed nucleus
and Hamiltonian, it acts on the internal structure of the HH representation and
decides how the resolution of the high-fidelity calculation is distributed;
no emulator of the solution is built.  The HH space is decomposed into
physically identifiable classes with independent hyperangular and
hyperradial cutoffs.  Starting from a compact space, a Gaussian-process (GP)
surrogate~\cite{Rasmussen2006} learns the marginal gain of each admissible
extension and uses its prediction and its uncertainty, weighted by
computational cost, to choose the next one.

A GP surrogate has recently been used to select basis functions in the
Gaussian expansion method, a nonorthogonal basis of correlated Gaussians for
few-body systems~\cite{Aoyama2026}.  There the surrogate chooses individual
functions; here it chooses the resolution of physically labelled sectors of an
orthonormal HH space, so what it returns is a truncation with quantum numbers.

 Every accepted
extension is followed by an explicit solution of the Hamiltonian in the
enlarged space, and the measured gain is returned to the learner, which
therefore learns where resolution is useful rather than learning the energy.
  Termination is built on the
same measurements: a coordinate is treated as exhausted when three
consecutive measured gains, taken after the hyperangular and hyperradial
resolutions of its class have settled with respect to each other, fall below
a threshold fixed by the target accuracy.  We also
examine what a posterior description of the unresolved space can deliver.

We test the approach on the $^3$H and $^3$He ground states, where
conventional HH calculations can be converged to benchmark accuracy and used
as references.  We study how the algorithm distributes resolution among
classes, quantify the reduction of basis dimension at fixed accuracy, assess
the residual truncation error, and test what one calculation can pass to
the next, between the two nuclei and between the two Hamiltonians.

The paper is organized as follows.  In Sec.~\ref{sec:hh} we summarize the
HH representation and its sector decomposition, in
Sec.~\ref{sec:active-learning} the adaptive construction, and in
Sec.~\ref{sec:results} the numerical setup and the results.  Our conclusions
are collected in Sec.~\ref{sec:conclusions}, while
Appendix~\ref{app:gp} specifies the Gaussian process and
Appendix~\ref{app:tail} the estimators of the unresolved tail.

\section{\label{sec:hh}Hyperspherical-harmonics framework}

\subsection{\label{sec:hh-basis}Basis for $A=3$}

We follow the HH formalism of Ref.~\cite{Marcucci2020} and write it here for
the three-nucleon systems studied in this work, with two conventions adapted
to the present implementation.  First, the Jacobi coordinates are ordered
pair-first, so that the relative coordinate of particles 1 and 2 is the first
Jacobi vector.  Second, instead of summing amplitudes over particle
permutations we work directly with an orthonormal basis of the fully
antisymmetric subspace. 
The intrinsic Hamiltonian is
\begin{equation}
 H = T + \sum_{i<j} V_{ij} + V_{123},
 \label{eq:hamiltonian}
\end{equation}
with $T$ the intrinsic kinetic energy, for which we take
$\hbar^2/m=41.47$~MeV\,fm$^2$, $V_{ij}$ the two-nucleon interaction,
which may include the electromagnetic terms, and $V_{123}$ the three-nucleon
interaction.  With the pair-first convention the two mass-scaled Jacobi
vectors are
\begin{equation}
 \bm{\eta}_1 = \frac{\bm{r}_1-\bm{r}_2}{\sqrt{2}},
 \qquad
 \bm{\eta}_2 = \sqrt{\frac{2}{3}}
 \left(\bm{r}_3-\frac{\bm{r}_1+\bm{r}_2}{2}\right),
 \label{eq:jacobi}
\end{equation}
which separate the center-of-mass motion and give
$T=-(\hbar^2/2m)(\nabla^2_{\bm{\eta}_1}+\nabla^2_{\bm{\eta}_2})$.  The
hyperradius $\rho^2=\eta_1^2+\eta_2^2$ is invariant under particle
permutations, and the remaining intrinsic variables are the two polar angles
$\hat{\bm{\eta}}_1,\hat{\bm{\eta}}_2$ together with a single hyperangle
$\phi$, defined by $\eta_1=\rho\sin\phi$ and $\eta_2=\rho\cos\phi$.  We write
$\Omega=\{\hat{\bm{\eta}}_1,\hat{\bm{\eta}}_2,\phi\}$.  In the $d=6$
dimensional intrinsic space
\begin{equation}
 T = -\frac{\hbar^2}{2m}
 \left[\frac{\partial^2}{\partial\rho^2}
 +\frac{5}{\rho}\frac{\partial}{\partial\rho}
 -\frac{\Lambda^2(\Omega)}{\rho^2}\right],
 \label{eq:kinetic-hh}
\end{equation}
with $\Lambda^2$ the grand-angular-momentum operator.  Its eigenfunctions are
the hyperspherical harmonics,
$\Lambda^2\mathcal{Y}=K(K+4)\mathcal{Y}$, given by
\begin{align}
 \mathcal{Y}_{KLM_L}^{l_1l_2n}(\Omega)
 &= \left[Y_{l_1}(\hat{\bm{\eta}}_1)Y_{l_2}(\hat{\bm{\eta}}_2)\right]_{LM_L}
 \mathcal{N}_n^{l_1l_2}
 \nonumber\\
 &\quad\times
 (\sin\phi)^{l_1}(\cos\phi)^{l_2}
 P_n^{(l_1+1/2,\,l_2+1/2)}(\cos2\phi),
 \label{eq:hh-A3}
\end{align}
with $P_n^{(a,b)}$ a Jacobi polynomial, $\mathcal{N}_n^{l_1l_2}$ a
normalization factor, and
\begin{equation}
 K = l_1+l_2+2n .
 \label{eq:K-A3}
\end{equation}
The pair-first convention associates the relative motion of particles 1 and
2 with $l_1$.

The spatial functions are coupled to spin and isospin states built starting
from the pair 1--2, $[[\chi_{1/2}(1)\chi_{1/2}(2)]_{s}\chi_{1/2}(3)]_{SM_S}$
with $s$ the pair spin, and likewise for isospin with pair isospin $t$.
Antisymmetry under exchange of particles 1 and 2 then requires
\begin{equation}
 l_1+s+t \quad \text{odd}.
 \label{eq:pair-antisymmetry}
\end{equation}

Complete antisymmetry under the exchange of the remaining pairs can be
imposed along three routes, all established in the literature: the
symmetrized HH construction, which sums amplitudes over particle
permutations and afterwards discards the redundant states that this
generates at fixed $K$~\cite{Kievsky2008,Marcucci2020,Gnech2020}; the
nonsymmetrized HH method, which forgoes symmetrization altogether and
recovers the permutational symmetry from the eigenvectors after
diagonalization~\cite{Gattobigio2011,Deflorian2013,Nannini2018}; and the
coefficient-of-fractional-parentage (CFP) construction, which builds
an orthonormal basis of the antisymmetric subspace directly, formulated for
HH functions in Refs.~\cite{Barnea1997,Barnea1998} and standard for
harmonic-oscillator Jacobi bases~\cite{Navratil2000}.  We adopt the last, for
two reasons that bear directly on the adaptive construction.

The states that satisfy Eq.~(\ref{eq:pair-antisymmetry}) and are coupled to
given $K$, $L$, $S$, $T$ and $J^\pi$ are antisymmetric under the exchange of
particles 1 and 2 only.  We call them parent states, $|p_\mu\rangle$, the
index $\mu$ running over the allowed values of $l_1$, $l_2$, $n$, $s$ and
$t$.  A completely antisymmetric state must also change sign under the
exchange $P_{23}$ of particles 2 and 3.  We therefore compute the matrix of
$P_{23}$ between parent states,
$\overline{P}_{\mu\nu}=\langle p_\mu|P_{23}|p_\nu\rangle$, and
diagonalize it.  Its eigenvectors with eigenvalue $-1$, whose components
$C_{\mu\alpha}$ are the coefficients of fractional parentage, give the
completely antisymmetric states, written as the kets
\begin{equation}
 |\mathcal{A}_{q\alpha}\rangle = \sum_\mu C_{\mu\alpha}|p_\mu\rangle ,
 \qquad
 \sum_\nu\overline{P}_{\mu\nu}C_{\nu\alpha} = -C_{\mu\alpha} .
 \label{eq:cfp}
\end{equation}
We call a sector the set of all independent antisymmetric states
$|\mathcal{A}_{q\alpha}\rangle$ with given $q=(K,L,2S,2T)$.  The label
$\alpha=1,2,\ldots$ numbers them, and their number is the rank of the
sector.

The first reason is conditioning.  With the core already antisymmetric,
$\overline{P}$ has the exact spectrum $\{-1,\,1/(A-1)\}$, the two irreducible
representations of $S_A$ whose restriction to $S_{A-1}$ is the sign
representation, so the antisymmetric states form an eigenspace set apart by a
gap $A/(A-1)$.  Extracting them is thus a well-separated eigenproblem, with
no near-degeneracy to resolve.  The difficulty this avoids is documented in
Ref.~\cite{Gnech2020}, where two classes had to be truncated above $K=10$ to
keep the permutation-sum route numerically stable.  Our implementation checks
the spectrum of $\overline{P}$ in every sector it builds.

The second reason is that the rank of each sector is known in advance, so
raising a cutoff has a predictable effect on the dimension.  Where selection
rules make that effect null the enlargement is skipped without a solve; were
it not, the measured gain of a move that adds nothing could not be told from
convergence.

For $A>3$ the same scheme applies with the pair replaced by an antisymmetric
$(A-1)$-body core coupled to the remaining nucleon, and $P_{23}$ by
$P_{A-1,A}$~\cite{Barnea1997,Navratil2000}.  The implementation stores the
CFPs by parentage label and never builds the $A$-body product space, so the
same construction runs unchanged at larger $A$.

The hyperradial dependence is expanded on a $K$-independent Laguerre basis,
so that
\begin{equation}
 \Psi^{JJ_z}_{M_T}
 = \rho^{-5/2}
 \sum_{q,\alpha}\sum_{m=0}^{N_{\rho,q}-1}
 c_{q\alpha m}\, f_m(\rho)\,
 |\mathcal{A}_{q\alpha}^{JJ_zM_T}\rangle ,
 \label{eq:wavefunction-expansion}
\end{equation}
\begin{equation}
 f_m(\rho)
 = \frac{1}{\sqrt{b}}\sqrt{\frac{m!}{\Gamma(m+6)}}\,
 z^{5/2}e^{-z/2}L_m^{(5)}(z),
 \label{eq:radial-basis}
\end{equation}
with $L_m^{(5)}$ a generalized Laguerre polynomial, $z=\rho/b$, and $b$ a
nonlinear length scale; the $f_m$ are orthonormal on $(0,\infty)$.  In
Eq.~(\ref{eq:wavefunction-expansion}) the $c_{q\alpha m}$ are the variational
coefficients and $N_{\rho,q}$ the number of radial functions retained in
sector $q$, which is what the hyperradial truncation controls.  Since both the radial and the antisymmetric HH states are
orthonormal, the Rayleigh--Ritz principle gives the ordinary eigenvalue
problem $\sum_j H_{ij}c_j=Ec_i$.  For nested spaces
$\mathcal{V}_1\subset\mathcal{V}_2$ one has
$E_0(\mathcal{V}_2)\leq E_0(\mathcal{V}_1)$; equivalently, with the binding
energy $B=-E_0$, enlarging the basis produces a non-negative gain.  This
monotonicity is the property the active-learning procedure exploits.

\subsection{\label{sec:hh-sectors}Sectors, classes and truncation}

The construction of the wave function decomposes the antisymmetric HH basis
on which it is expanded into the sectors
$q=(K,L,2S,2T)$ of Eq.~(\ref{eq:cfp}).  A sector differs from the traditional HH
``channel'', which fixes in addition the individual quantum numbers $l_1$,
$l_2$, $n$, $s$ and $t$.  In a sector these labels are combined, through the
coefficients $C_{\mu\alpha}$ of Eq.~(\ref{eq:cfp}), into the index $\alpha$,
which only numbers its independent states.  Sectors that
differ only in $K$ are grouped into classes
\begin{equation}
 c=(L,2S,2T)\in\mathcal{C}.
 \label{eq:class-definition}
\end{equation}
Parity restricts $K$ through $(-1)^K=\pi$.

A conventional uniform truncation retains every sector with $K\leq K_{\max}$
and gives every sector the same number $N_\rho$ of radial functions: two
numbers fix the space.  The adaptive construction of
Sec.~\ref{sec:active-learning} assigns one hyperangular cutoff $K_c$ to each
class, so that sector $q$ is retained whenever $K\leq K_c$ for its class $c$,
and treats the hyperradial direction in one of two modes, both acting on
classes: a single $N_\rho$ for the whole space, the uniform-radial
mode, or one cutoff $N_{\rho,c}$ per class, shared by all the sectors of that
class, the class-radial mode.  The truncated space is
thus
\begin{equation}
 \mathcal{V}(\bm{K},\bm{N}_\rho)
 = {\rm span}\left\{|q\alpha m\rangle:\,
 K\leq K_c,\; m<N_{\rho,c}\right\},
 \label{eq:truncated-space}
\end{equation}
with $\bm{K}=(K_{c_1},K_{c_2},\ldots)$ and
$\bm{N}_\rho=(N_{\rho,c_1},N_{\rho,c_2},\ldots)$; the uniform truncation is
the special case $K_c=K_{\max}$ and $N_{\rho,c}=N_\rho$ for all $c$.  The
adaptive construction selects both vectors from the numerical information
accumulated during the calculation, leaving the variational problem
unchanged.

\section{\label{sec:active-learning}Active-learning construction}

\subsection{\label{sec:adaptive-coordinates}Adaptive coordinates and marginal
gains}

The adaptive coordinates are the cutoffs of Eq.~(\ref{eq:truncated-space}) in
Sec.~\ref{sec:hh-sectors}.  An elementary angular refinement raises the
hyperangular cutoff of one class, $K_c\rightarrow K_c+2$, and an elementary
radial refinement its number of Laguerre functions,
$N_{\rho,c}\rightarrow N_{\rho,c}+\Delta N_\rho$, with $\Delta N_\rho$ a
step fixed for the run.  We call shell the set of
states that one elementary refinement adds to a class: the sector with the new
value of $K$, or the new Laguerre functions of all its sectors.  The angular coordinates are labelled
by the classes $\mathcal{C}$ of Eq.~(\ref{eq:class-definition}), and we
denote by $\mathcal{R}$ the set of radial coordinates, one per class.
Resolving the hyperradial direction per class adds a further reduction, since
a class whose angular content saturates early also saturates early in $\rho$,
as shown below in Sec.~\ref{sec:results-adaptive}.

Writing $\mathcal{I}=\mathcal{C}\cup\mathcal{R}$ for the complete set of
adaptive coordinates, the state of the calculation is
\begin{equation}
 \bm{x} = \left(\{K_c\}_{c\in\mathcal{C}};
 \{N_{\rho,c}\}_{c\in\mathcal{C}}\right),
 \label{eq:adaptive-state}
\end{equation}
and $\bm{x}\rightarrow\bm{x}+\delta_a$ denotes the elementary refinement of
coordinate $a\in\mathcal{I}$.  Every rank-increasing refinement adds states
without removing any, so the spaces visited along a trajectory are nested.

The learner models the marginal gain of a candidate refinement,
\begin{equation}
 g_a(\bm{x}) = B(\bm{x}+\delta_a)-B(\bm{x}) \geq 0 ,
 \label{eq:marginal-gain}
\end{equation}
non-negative up to the accuracy of the eigensolver.  Gains span several
orders of magnitude during a run, so the modeled variable is
\begin{equation}
 y_a(\bm{x}) = \log\left[g_a(\bm{x})+G_0\right],
 \qquad G_0=10^{-6}~{\rm MeV},
 \label{eq:log-gain}
\end{equation}
with $G_0$ a small offset that keeps the transformation defined for
numerically vanishing gains.

\subsection{\label{sec:gp-surrogate}Gaussian-process surrogate}

Evaluating Eq.~(\ref{eq:marginal-gain}) for every candidate at every
iteration would require exactly the many solves the procedure is meant to
avoid.  We therefore model the gain field with a GP~\cite{Rasmussen2006}.

Each adaptive coordinate carries a refinement index $n_a$: zero when
inactive, one at its first active value, and one more for each further
shell.  The GP input is the full vector
$\bm{n}=(n_{a_1},\ldots,n_{a_{N_a}})$ of the $N_a$ coordinates, so the
predicted gain of one coordinate depends on the resolution of the whole
space.  The transformed gains are modeled as
$y_a(\bm{x})\sim\mathcal{N}[\mu_a(\bm{x}),\sigma_a^2(\bm{x})]$, with
$\mu_a$ and $\sigma_a^2$ the posterior mean and variance, and with a prior
mean
\begin{equation}
 m_a(n_a) = \alpha_a-\beta_a n_a
 \label{eq:gp-mean}
\end{equation}
that encodes an approximately exponential decay of successive contributions.
The local parameters $(\alpha_a,\beta_a)$ are
fitted to the gains measured for coordinate $a$; before enough local data exist they come from an
optimistic default or from a transferred prior, discussed in
Sec.~\ref{sec:transfer-method}.

Coordinates are correlated through the kernel
\begin{equation}
 k\!\left[(\bm{n},a),(\bm{n}',a')\right]
 = e^{-\|\bm{n}-\bm{n}'\|^2/2\ell^2}
 \left[\sigma_{\rm sh}^2+\delta_{aa'}\sigma_{\rm ind}^2\right].
 \label{eq:gp-kernel}
\end{equation}
Two observations are correlated when the spaces they were made in are
close, and $\ell$ is the length scale over which that correlation decays.
The shared amplitude $\sigma_{\rm sh}$ lets observations made while refining
one coordinate inform the others; the individual amplitude $\sigma_{\rm ind}$,
which enters only when $a=a'$, retains the behavior specific to each
refinement direction.  Hyperparameters are inferred from the accumulated data by regularized
marginal-likelihood optimization, and the predictive covariance is calibrated
on leave-one-out errors, a calibration that can enlarge but never shrink the
nominal GP uncertainty.  Appendix~\ref{app:gp} collects the corresponding
expressions.

No precomputed training set is needed.  Starting from an initial space, the
Hamiltonian is solved, the GP predicts the gain of every admissible
refinement, one is selected, the Hamiltonian is solved in the enlarged space,
and the measured gain is added to the training set.  Repeating this cycle
learns the convergence pattern of the nucleus, Hamiltonian and
representation at hand.

\subsection{\label{sec:acquisition}Cost-aware acquisition}

To choose the next refinement, the learner weighs three quantities: the
predicted gain, its uncertainty and the cost of the solve.  We fix a small
gain threshold $g_{\rm tol}$, below which a refinement is not worth a solve.
For each candidate $a$ we then compute the expected amount by which its gain
exceeds the threshold,
\begin{equation}
 U_a(\bm{x}) = \mathbb{E}\left[\max\left(g_a-g_{\rm tol},0\right)\right],
 \label{eq:expected-excess}
\end{equation}
where $\mathbb{E}$ denotes the expectation value, i.e.\ the average over the
GP prediction of $g_a$.  $U_a$ is therefore large
when a large gain is predicted, but also when the gain is uncertain and may
well exceed $g_{\rm tol}$.  This is the expected-improvement criterion of
Bayesian optimization~\cite{Jones1998}, with the best value found so far
replaced by the fixed threshold; for the lognormal prediction of
Eq.~(\ref{eq:log-gain}) it has a closed form.

The cost of a candidate space is known before it is solved, through the
retained dimension $N$ and the number of retained sector pairs
$N_{\rm pair}$, both fixed by the resolution vector $\bm{x}$.  We model the
cost $C(\bm{x})$ as a low-order polynomial in these two quantities,
\begin{equation}
 C(\bm{x}) = N(\bm{x}) + \left[\frac{N(\bm{x})}{N_\times}\right]^{3}\! N_\times
 + c_{\rm b}\,N_{\rm pair}(\bm{x}) ,
 \label{eq:cost-model}
\end{equation}
where the first two terms are an empirical description of the eigensolver
cost, linear at small dimension and cubic beyond a fitted crossover
$N_\times$, and the third the
Hamiltonian construction, which scales with the number of sector-pair blocks
rather than with the basis dimension.  The coefficients were fitted once,
offline, to timed solves, giving $N_\times\simeq2300$ and $c_{\rm b}=62$ in
units of the linear term, with no term in $N^2$; they are then held fixed.  The
acquisition uses the predicted
increment $\Delta C_a=C(\bm{x}+\delta_a)-C(\bm{x})$, and the score is
\begin{equation}
 S_a(\bm{x}) = \frac{U_a(\bm{x})}{\left[C_a^{\rm acq}(\bm{x})\right]^p},
 \qquad
 a^\star=\arg\max_{a\in\mathcal{I}_{\rm adm}} S_a ,
 \label{eq:acquisition-score}
\end{equation}
where $C_a^{\rm acq}=\max(\Delta C_a,C_{\min})$ is that increment floored by
$C_{\min}=10^{-6}\,C(\bm{x})$, so the denominator cannot vanish on a
rank-neutral move, $p=1$ in the production runs,
$\mathcal{I}_{\rm adm}$ is the set of admissible refinements, and $a^\star$,
the refinement with the largest score, is the one selected.  The incremental cost enters only this ranking; each selected
rank-increasing candidate is still evaluated by a complete solution of the
Hamiltonian in the enlarged space.

Two safeguards complete the selection.  The shared covariance in
Eq.~(\ref{eq:gp-kernel}) can make an untested direction look irrelevant
before it has ever been probed, so we require a minimum number of
informative rank-increasing observations, $n_{\rm min}$, in a designated
exploration set $\mathcal{I}_{\rm exp}\subseteq\mathcal{I}$ before
posterior-based stopping is allowed; the production runs take
$n_{\rm min}=1$ and $\mathcal{I}_{\rm exp}=\mathcal{I}$, so every coordinate
must be probed at least once.
Second, a formal increase of a cutoff need not enlarge the space.  Such
rank-neutral actions are detected before any diagonalization: the coordinate
advances, and a zero gain is recorded as a soft-censored observation,
one with a much larger observation variance, weak evidence for a small gain that
does not suppress the coordinate permanently.  The first rank-increasing
probe of a coordinate is treated the same way if its gain is numerically
negligible, giving it one chance to reawaken; later small gains count as
ordinary observations.

\subsection{\label{sec:stopping}Termination}

A run stops when it has established in two independent ways that the binding
energy still available is below the requested accuracy $\epsilon_E$: from
what it has measured, and from what it predicts about the refinements it has
not computed.  The measured tests see only the directions the trajectory has
visited; the predictions cover the rest, but are loose.

The measured tests are three.  First, every coordinate $a$ that can still be
grown, that is with current value $x_a$ below its largest value $x_a^{\max}$
($K_{\max}$ for a hyperangular cutoff, $N_\rho$ for a hyperradial one), must
have shown that growing it buys less than $\epsilon_E$, in three consecutive
current readings.  A reading is current until the other resolution of its
class changes what it measured: a hyperradial reading is retired once the
hyperangular cutoff of its class has grown.  Hyperangular readings need no
such rule, because every class enters with six Laguerre functions; read with
fewer, its angular gain would not be the one the developed class shows, as
shown in Sec.~\ref{sec:setup-protocol}.  The rule is internal to a class:
readings are not retired when a different class grows, as discussed in
Sec.~\ref{sec:results-uq}.  There is one exception: every reading taken while
the lowest eigenvalue lies above threshold is retired when the state binds.
Until then the lowest level is a discretized continuum state, fixed by the
hyperradial basis, and what a class adds to it says nothing of what the class
will add to the bound state.  A single sub-tolerance reading is not evidence
either, since along a Laguerre ladder a silent shell can be followed by one
carrying real strength.  Second, the last measured gains $g_a^{\rm last}$ of
those coordinates must sum to less than $\epsilon_E$, i.e.
\begin{equation}
 S_{\rm front} = \sum_{a\,:\,x_a<x_a^{\max}} g_a^{\rm last}
 \leq \epsilon_E .
 \label{eq:frontier-sum}
\end{equation}
This test is needed because the coordinates can pass the first one
individually while their sum does not: eight coordinates gaining $4$~keV each
pass one by one, yet together they would still add $32$~keV.  Third, the last
two computed gains must lie below $\epsilon_E$.

The predicted tests are two.  First of all, no single admissible refinement
may still be worth taking: the largest upper-confidence-bound gain
$g_a^{\rm UCB}=\max[e^{\mu_a+\kappa\bar{\sigma}_a}-G_0,0]$ must fall below
$\epsilon_E$.  Here $\bar{\sigma}_a$ is the calibrated posterior width, as
reported in Appendix~\ref{app:gp}, and $\kappa$ the optimism of the bound, as
in the upper-confidence-bound rules of Gaussian-process
optimization~\cite{Srinivas2012}.  Second, the gains still available up to the
largest space the run may reach, $K_{\max}$ for the hyperangular coordinates
and $N_\rho$ for the hyperradial ones, must sum to less than $\epsilon_E$.
That sum, $\Delta_{K_{\max}}$, is the median of three estimates: the GP
extrapolated jointly up to $K_{\max}$ and $N_\rho$, a decay law fitted to each
coordinate, and a nonparametric envelope built from the gain ratios observed
along the run.  Its spread $\sigma_{K_{\max}}$ is taken from their dispersion
and from the GP posterior; both are given in Appendix~\ref{app:tail}.
Nothing guarantees that the estimate is conservative, so it is calibrated
within the run, in a construction modelled on conformal
prediction~\cite{Lei2018} without its distribution-free guarantee, which
assumes exchangeable data.  Each earlier prediction is compared with the
binding actually gained since, and the $\gamma$ quantile of those ratios,
floored at one, gives a scale $s_{\rm conf}$ that may inflate the estimate but
never shrink it.  The certified residual is
\begin{equation}
 \Delta_{\rm cert} = s_{\rm conf}
 \left(\Delta_{K_{\max}}+z\,\sigma_{K_{\max}}\right) ,
 \qquad z=\Phi^{-1}\!\left[\tfrac{1+\gamma}{2}\right],
 \label{eq:certified-bound}
\end{equation}
with $\Phi$ the standard normal distribution function and $\gamma$ the
calibration level, and this test requires $\Delta_{\rm cert}<\epsilon_E$.
$\Delta_{\rm cert}$ is the estimate of the run of what the $K_{\max}$ space still
holds beyond the selected one, and whether it covers it is tested on the
benchmarks of Sec.~\ref{sec:results-uq}.  What lies beyond $K_{\max}$ is
outside its reach.

A run that satisfies both families of tests is certified: its accuracy then
rests on a calibrated estimate.  A run may also end because no admissible
refinement is worth more than the per-move threshold $g_{\rm tol}$, or because
the step budget runs out.  The two families of tests are then reported
separately, so that a run whose measured evidence is complete, and whose
extrapolated estimate is only too cautious to certify, can be told from one
that stopped short.  The first is converged on what it measures, which is how
the observable-targeted runs discussed in Sec.~\ref{sec:results-observables}
end.

The rule never refers to an experimental value, whose distance from the
computed energy contains the error of the Hamiltonian, some $860$~keV on the
$^3$H binding energy for the Argonne $v_{18}$ interaction without a
three-nucleon force.  Where a uniform
ladder is available, the difference against its largest space is recorded as
an independent check, discussed in Sec.~\ref{sec:results-adaptive}, and never
used to stop.

A natural refinement is to ask the Hamiltonian rather than the surrogate.
Given the current eigenvector $|\psi\rangle$, the expansion of
Eq.~(\ref{eq:wavefunction-expansion}) in the selected space, second-order
perturbation theory on the excluded space can estimate the energy it still
holds, calculating the matrix elements $\langle j|H|\psi\rangle$ between
$|\psi\rangle$ and the basis states $|j\rangle$ of the sectors left out, in
the spirit of the
importance-truncation schemes of Refs.~\cite{Roth2007,Roth2009}.  It costs
one matrix--vector product per check and covers directions the trajectory
never explored, at the price of touching an excluded space that at high $K$
outweighs the retained one in memory and time.  We return to this point in
Sec.~\ref{sec:conclusions}.

\subsection{\label{sec:observable-targets}Observable targets}

Convergence of the energy does not imply convergence of other quantities, so
scalar observables can be added as targets.  For each observable $O$ a
separate gain-field GP models the absolute change
\begin{equation}
 g_a^{(O)}(\bm{x}) = \left|O(\bm{x}+\delta_a)-O(\bm{x})\right| .
 \label{eq:observable-gain}
\end{equation}
This quantity is not variational and the signed change need not be
monotonic; its absolute value is a conservative measure of the accumulated
representation dependence.  With $\epsilon_O$ the requested tolerance on $O$,
changes are mapped onto the energy scale by
$\widetilde g_a^{(O)}=(\epsilon_E/\epsilon_O)\,g_a^{(O)}$, so that
$g_a^{(O)}=\epsilon_O$ corresponds to $\widetilde g_a^{(O)}=\epsilon_E$.  The
acquisition utilities are then summed after rescaling, in units of
$g_{\rm tol}$, and Eq.~(\ref{eq:acquisition-score}) applies with $U_a$
replaced by that combined utility.  The targets enter the numerator of the
score and not its denominator: the cost of a solve does not depend on what
is then measured in the space it defines.  A refinement can therefore be selected for
its effect on an observable even when its energetic effect is small.  Every target must meet
its own conditions, in the same three forms as the energy, before the run
is certified.

\subsection{\label{sec:transfer-method}Transfer between calculations}

Different calculations often converge in the same way.  Mirror nuclei have
the same classes and, apart from Coulomb and charge-symmetry-breaking
effects, the same class content.  Similarly, two states of one nucleus, or
one nucleus described by two Hamiltonians, share the correlations that
dominate them; the tensor-induced $D$ state of the deuteron, for instance,
reappears in every nucleus that contains a spin-one, isospin-zero nucleon
pair.  In these cases a second calculation need not rediscover from scratch
the order in which a first one found resolution to be useful: what one
adaptive calculation has learned can initialize another, saving the
exploratory solves of a cold start.  We use two forms of transfer, which pass
different information and act at different moments.  Neither passes energies
or wave functions.

The first form transfers the decay law, i.e. how the gains of the source run
decayed along each coordinate, Eq.~(\ref{eq:gp-mean}).  Its intercept is
taken relative to the largest gain of the source run, and the optimized
hyperparameters of Appendix~\ref{app:gp} are transferred as well.  This form
only initializes the surrogate: as soon as a coordinate of the target has two
informative readings of its own, its parameters are fitted to those readings
alone.  It can change the order in which the target representation is
explored and, through it, slightly, the space at which the run stops and the
number of solves.

The second form transfers the resolution vector, i.e.\ essentially where the
source run ended.  With $\bm{x}^\ast$ the final state of the source, in the
notation of Eq.~(\ref{eq:adaptive-state}), the target run starts from
$\bm{x}^\ast$, taking every coordinate three shells back, never below its
minimum and never above $K_{\max}$, instead of from the smallest space.  From
there the learner proceeds as in a cold start, with an untrained surrogate.
Three shells are what the measured tests ask of a coordinate before it
can be declared spent.  In this way, a target that needs no more resolution
than the source is certified after three readings per coordinate, while one
that needs more goes on growing where the readings say so.  The vector only
sets the starting point, and every coordinate stays open up to $K_{\max}$.
This transfer removes the path from the smallest space to the neighborhood
of the final one, but the three readings per coordinate that certify the
final space are still taken.

\section{\label{sec:results}Results}

\subsection{\label{sec:setup-systems}Systems, Hamiltonians and basis}

We compute the $J^\pi=\frac12^+$ ground states of $^3$H and $^3$He, using
$T_z=-1/2$ and $+1/2$ respectively and retaining both $T=1/2$ and $T=3/2$ to
keep the small isospin admixtures generated by the charge-dependent
interaction.  Two Hamiltonians are used: the Argonne $v_{18}$ two-nucleon
interaction (AV18)~\cite{Wiringa1995} with its complete electromagnetic
part, and AV18 plus the Urbana IX three-nucleon interaction
(UIX)~\cite{Pudliner1995,Pieper2001} with the parameters of
Ref.~\cite{Pudliner1995}, retaining both its two-pion-exchange component
$V_{2\pi}$ and its short-range component $V_R$.  The two Hamiltonians are
diagonalized in the same HH basis, with the same cutoffs and radial
functions, so any difference between the spaces the learner selects with and
without UIX is due to the three-nucleon interaction alone.  The
implementation can also handle the nonlocal interactions derived from chiral
effective field theory~\cite{Epelbaum2009,Machleidt2011,Entem2017}.
However, we limit the present study to AV18 and AV18+UIX, local interactions
with a strong short-range repulsion, which are numerically more demanding in
the HH expansion (see, for instance, Ref.~\cite{Gnech2020} for $^6$Li).

For $J^\pi=1/2^+$ the angular-momentum couplings admit $L\leq2$ only, and
leave eight nonvanishing classes $c=(L,2S,2T)$.
The hyperradial expansion uses at most $N_\rho=28$ Laguerre functions per
sector with length parameter $b=0.20$~fm in Eq.~(\ref{eq:radial-basis}),
chosen from a variational scan for $^3$H with AV18: for $b$ from $0.20$ to
$0.40$~fm it gave the lowest energy at every sufficiently developed hyperangular
truncation, and the neighboring value $b=0.25$~fm differed by about
$0.1$~keV.  With this radial basis the converged ladders reproduce the
published HH benchmarks to within $0.9$~keV, as discussed in
Sec.~\ref{sec:results-reference}.
The values of $N_\rho$ and $b$ are held fixed in all reference calculations.  In the
adaptive runs $b$ is the same, and the number of Laguerre functions is
learned only in the class-radial mode, where each class retains
$N_{\rho,c}\leq N_\rho$ of them; in the uniform-radial mode every sector
keeps all $N_\rho$.

All angular integrals are computed with Gauss quadratures whose number of
integration points grows with the grand angular momenta $K$ and $K'$
of the bra and ket sectors of each matrix block.  The hyperangle of the
two-nucleon matrix elements uses $K_{\max}+16$ integration points, rounded up
to a
multiple of $20$, i.e.\ $80$ for $K_{\max}=60$; the hyperangle of the
three-nucleon ones uses $\max[24,(K+K')/2+2]$ integration points, and the
angle between
the two Jacobi vectors $\max[32,(K+K')/2+20]$.  Quadratures of fixed order,
adequate at small $K$, are not adequate at large $K$: with twelve integration
points on the
Jacobi angle the three-nucleon blocks at $K$ from $40$ to $60$ are wrong by
up to $25$\%.  With the rules used here, doubling them changes
the largest blocks at $K=60$ by less than one part in $10^4$ and the
AV18+UIX energies at $K=20$ and $K=28$ by at most $0.004$~keV.  The
hyperradial integrals use $150$ Gauss--Laguerre integration points, and the
energies
agree to $10^{-3}$~keV with those obtained with $300$.

Spaces below $1200$ states are diagonalized directly; in larger ones the
lowest eigenvalue is obtained with the Lanczos method~\cite{Lanczos1950} as
implemented in ARPACK~\cite{Lehoucq1998}, converged to a relative tolerance
of $10^{-7}$, which at these binding energies is below $1$~eV and therefore far
below the accuracy targets quoted here.  For every exact solve we record the
retained dimension and the numbers of active sectors and sector pairs, which
are the quantities the cost model of Eq.~(\ref{eq:cost-model}) is built from.

\subsection{\label{sec:setup-protocol}Reference ladder and adaptive settings}

For each nucleus and Hamiltonian we first build a reference sequence, the
conventional uniform truncation: all eight classes grow together, in steps of
$\Delta K=2$ up to $K=40$ and of $\Delta K=4$ from there to $K_{\max}=60$.
A truncation at $K$ holds $N_\rho(K+1)(K+2)/2$ states, so the last space of
the sequence has $N_{\rm full}=52\,948$.  We call the sequence the
reference ladder, and $E_{\rm last}$ the energy of its last space.  It
provides the baseline for the adaptive comparison, measures the residual
convergence of the conventional expansion, and defines the largest space
the learner may reach.  Where the last increments decrease geometrically,
the remaining tail $\Delta_{\rm geo}$ is obtained by summing the
corresponding geometric series, and
$E_{\rm ref}=E_{\rm last}-\Delta_{\rm geo}$ is the extrapolated reference
energy.

Adaptive runs start from the smallest nonvanishing angular sectors and, in
the class-radial mode, from $N_{\rho,c}=6$, increasing in steps of $\Delta N_\rho=2$ up to
$N_\rho=28$; angular coordinates increase in steps of $\Delta K=2$ up to
$K_{\max}=60$ for every class.  The starting value is a measured choice.
Started from two Laguerre functions, a class is first read with a radial
basis too poor for its hyperangular gain to mean anything, and the run sits
on plateaus of solves that gain nothing: on $^3$H with AV18, under the same
stopping criterion, starting from six takes $156$ exact solves instead of
$174$, of which $61$ instead of $81$ gain less than a hundredth of the
error then remaining, and certifies a space $17$\% smaller; it comes
within $20$~keV of the $K_{\max}$ space with $6398$ states instead of
$9520$.  The production runs use the class-radial mode; one companion run in
the uniform-radial mode, every sector keeping the full $N_\rho=28$, isolates
how much of the reduction comes from the hyperradial direction.

The accuracy requested of every adaptive run is $\epsilon_E=5$~keV on the
energy; $\epsilon_E$ denotes this value throughout.  The choice reflects that independent solutions of the
same trinucleon Hamiltonian agree at the level of a few
keV~\cite{Nogga2003,Kamada2001} and the published HH values used as a check
carry a $1$~keV uncertainty~\cite{Marcucci2020}, while the error of the
interaction itself is larger by one to two orders of magnitude, as seen in
Sec.~\ref{sec:results-observables}.  For the other parameters of the construction we use the following values.
The predicted tests of Sec.~\ref{sec:stopping} use the calibration level
$\gamma=0.90$ in Eq.~(\ref{eq:certified-bound}) and the optimism $\kappa=0.5$
in the upper-confidence-bound gain.  The acquisition of
Eq.~(\ref{eq:expected-excess}) considers a refinement worth a solve only if
its gain may exceed $g_{\rm tol}=0.01$~keV.  In the measured tests, a
coordinate is declared spent after three current readings below
$\epsilon_E$, and the last two computed gains must also lie below it;
readings taken before the state is bound are retired when it binds.  Each
move adds a single shell, and a run that needs more than $n_{\max}=4000$
exact solves is recorded as truncated rather than converged.

\subsection{\label{sec:setup-diagnostics}Diagnostics and comparison metrics}

We characterize each final state in two ways.  The probability carried by
class $c$ is $W_c=\sum_{q\alpha m\in c}|c_{q\alpha m}|^2$, the sum of the
squared coefficients of Eq.~(\ref{eq:wavefunction-expansion}) over all its
sectors and retained radial functions.  Its energetic importance is measured by an
explicit removal test: all sectors of the class are deleted, the Hamiltonian
is rediagonalized, and
\begin{equation}
 \Delta E_c = E_0(\mathcal{V}\setminus c)-E_0(\mathcal{V}),
 \label{eq:class-removal}
\end{equation}
with $\mathcal{V}\setminus c$ the space $\mathcal{V}$ without the sectors of
class $c$ and $E_0$ the lowest eigenvalue.  Since these are new variational
calculations, $\Delta E_c$ measures the true
energetic cost of the class, including the relaxation of everything else.
Consequently, the $\Delta E_c$ are not additive.  Removing only the
sector with the largest $K$ of a class, i.e.\ lowering its cutoff from $K_c$ to
$K_c-2$, gives the analogous saturation diagnostic
$\Delta E_c^{\rm top}$.  The same removals are used for the one-body
observables: the matter, point-proton, point-neutron and charge rms radii,
$r_m$, $r_p$, $r_n$ and $r_{\rm ch}$, and the magnetic moment $\mu$,
evaluated in the one-body impulse approximation.  The
spectroscopic quadrupole moment vanishes identically for $J=1/2$.  The
observable-targeted runs require a relative accuracy of $2\times10^{-3}$ on
every nonvanishing scalar observable in addition to the energy criterion:
$0.0035$~fm on the charge radius and $0.005$~$\mu_N$ on the magnetic moment,
below the experimental uncertainties of the charge radii~\cite{Amroun1994}
and well below the effects the calculation is meant to expose, the three-nucleon shift of the
radius and the two-body currents absent from the one-body magnetic moment
calculation~\cite{Marcucci2005,Marcucci2016}.

To compare representations at equal accuracy we measure the error of a
space from $E_{\rm last}$, the energy of the uniform $K_{\max}$ space.  For an
accuracy $\epsilon$ we take the smallest space of the adaptive run whose
energy lies within $\epsilon$ of $E_{\rm last}$, of dimension
$N_{\rm ad}(\epsilon)$, and compare it with the largest uniform space of the
ladder, of $N_{\rm full}=52\,948$ states.  Therefore we define
\begin{equation}
 R = 1-\frac{N_{\rm ad}}{N_{\rm full}}
 \label{eq:basis-reduction}
\end{equation}
as the fraction of the $K_{\max}$ space that need not be built to reach the
stated accuracy.  All reductions quoted below use this denominator.  There is one adaptive run per system, asked for
$\epsilon_E$: the values of $N_{\rm ad}(\epsilon)$ are read along its path,
for $\epsilon=\epsilon_E$ as for the looser accuracies, and are not the
outcome of runs asked for each $\epsilon$.  A uniform
calculation asked for a looser accuracy would stop earlier as well, so we
also give $N_{\rm unif}(\epsilon)$, the first space of the ladder within
$\epsilon$ of $E_{\rm last}$, and the reduction against it,
$R'=1-N_{\rm ad}/N_{\rm unif}$.
The cost of a representation is reported as the number of exact solves it
took to find, together with the time spent in exact solves where it is
relevant.  All production runs were executed on identical nodes of the
Leonardo system at CINECA~\cite{Leonardo}.

Finally, transfer is tested in three cases, each made of three runs: $^3$He
from
$^3$H, with each Hamiltonian, and $^3$H with AV18+UIX from $^3$H with AV18
alone, the case of a three-nucleon interaction added to a nucleus already
learned.  Each target is run three times: from a cold start, with the
surrogate initialized by the decay laws of the source, and from the
resolution vector of the source, see Sec.~\ref{sec:transfer-method}.  Energy
tolerance, $K_{\max}$ and stopping rule are identical for all the three runs.

The learner selects without knowing the answer.  To measure what that
costs we compare it with a greedy oracle, a construction that makes the same
kind of choice with knowledge the learner does not have.  It uses the
same HH code and grows the same space through the same moves as the
learner: sixteen coordinates, one hyperangular cutoff and one hyperradial
count per class, grown one shell at a time under the same ceiling
$K_{\max}=60$.  The two approaches differ in how the next move is chosen.  The learner
estimates the gain of every admissible move from its surrogate and solves
only the one it selects.  The greedy oracle solves all of them exactly, reads off
the true gain of each, and keeps the move with the largest energy gain per
state added; it then repeats from the enlarged space, and stops when its
error against $E_{\rm ref}$ falls below $\epsilon_E$.  It pays one solve
per admissible move per step, while the learner pays one per step, so it
serves only as a term of comparison.  The search is greedy: it takes the best
move at each step without looking further ahead, so the space it reaches
need not be the smallest possible one.  We run
it for $^3$H with both Hamiltonians, AV18 and AV18+UIX.

The greedy oracle is the term of comparison that knows more than the learner.  The
one that knows less is a rule that chooses without a surrogate: a
coordinate is worth the gain it gave the last time it was grown, divided by
the cost of growing it again, and a coordinate never grown is worth the
largest gain seen so far, so that it is read before it is ranked.
Everything else is as in the learner, the termination tests of
Sec.~\ref{sec:stopping} included, so that the two runs differ only in how
the next move is chosen.  We run it for $^3$H with AV18.

\begin{figure}[t]
\centering
\includegraphics[width=0.95\linewidth]{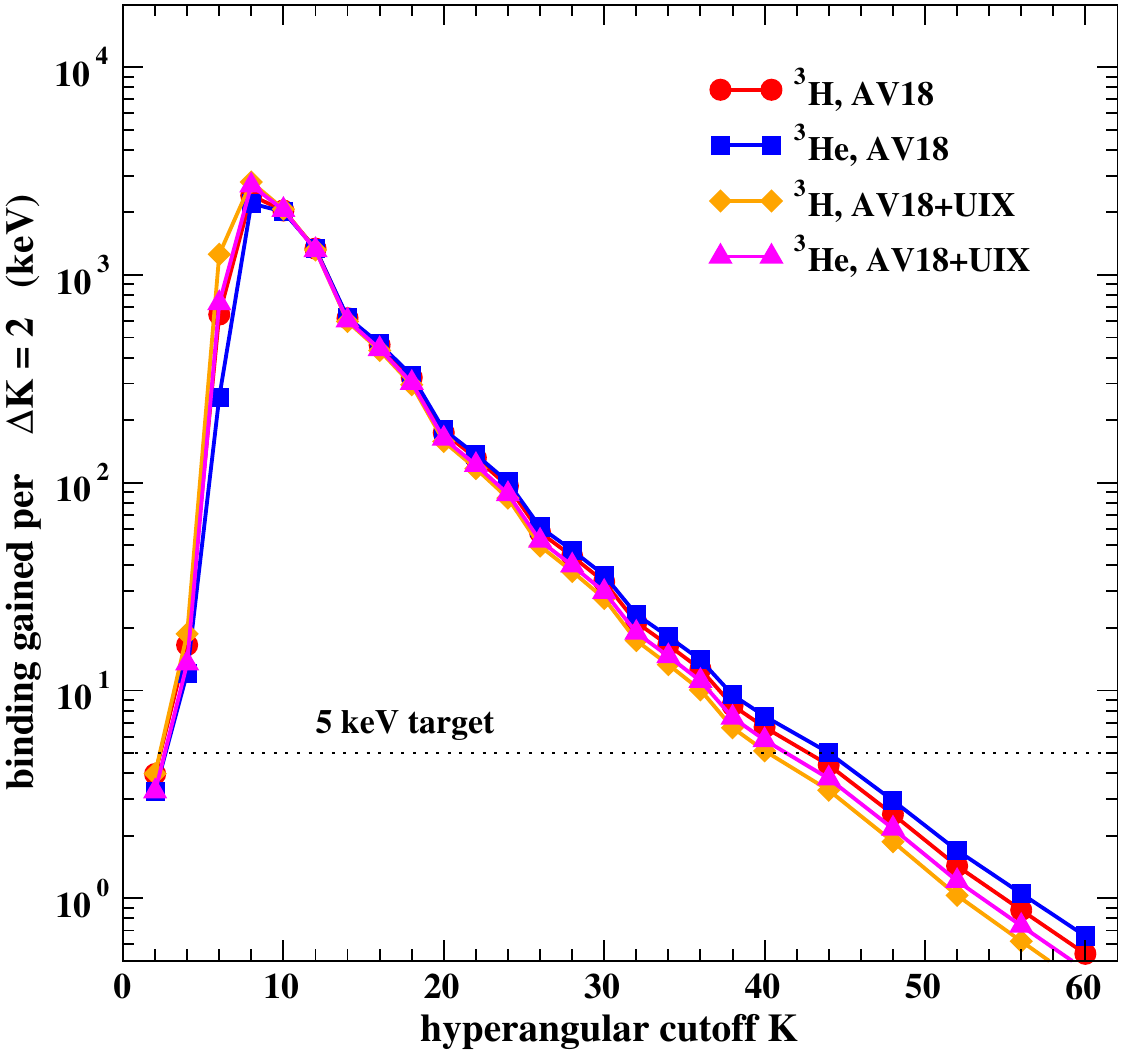}
\caption{\label{fig:reference-convergence}
Binding energy gained per $\Delta K=2$ along the uniform ladder, against the
hyperangular cutoff $K$; above $K=40$ the ladder moves in steps of four, and
the gain of each step is divided by two.  The increments are still above the
$5$~keV target at $K=40$ and decrease geometrically thereafter, which is
what allows the remaining tail to be summed (see
Table~\ref{tab:reference-energies} and the discussion in the text).}
\end{figure}

\subsection{\label{sec:results-reference}Reference convergence}

What each step of the conventional ladder is still worth is shown in
Fig.~\ref{fig:reference-convergence} for the four nucleus--Hamiltonian
combinations.  Every increment is positive, as required by the variational
principle, and after a maximum near $K=8$ the sequence falls by four orders of
magnitude.  In the largest space, $K_{\max}=60$ and $52\,948$ states, we
obtain $E_{\rm last}=-7.62138$, $-6.92285$, $-8.47836$ and $-7.74832$~MeV for
$^3$H and $^3$He with AV18 and with AV18+UIX.

The final increments decrease geometrically---the straight segment beyond
$K=44$ in Fig.~\ref{fig:reference-convergence}---so the remaining
hyperangular tail can be summed as a geometric series, giving residuals
$\Delta_{\rm geo}$ from $1.2$~keV ($^3$H, AV18+UIX) to $2.2$~keV ($^3$He, AV18)
and the extrapolated
references $E_{\rm ref}$ of Table~\ref{tab:reference-energies}.  The
binding accumulated between $K=40$ and $K=60$ is $19.5$ and $22.8$~keV for
$^3$H and $^3$He with AV18, and $14.4$ and $16.7$~keV with AV18+UIX, three to
four times the target.  At $K=60$ every class is
saturated: lowering the cutoff of a class from $K=60$ to $K=58$ in the final
space shifts the energy by less than $0.5$~keV for the $S$- and $D$-wave $T=1/2$
classes
and by less than $10^{-3}$~keV for every other.

The corresponding binding energies differ from the established HH
benchmarks~\cite{Marcucci2020} by $-0.9$ and $0.0$~keV for $^3$H and $^3$He
with AV18 and by $+0.5$ and $-0.3$~keV with AV18+UIX.  All four therefore
reproduce the published values within the $5$-keV target adopted here, and
within the $1$-keV uncertainty quoted for the benchmarks themselves.

\begin{table}[t]
\caption{\label{tab:reference-energies}
Reference energies.  $E_{\rm last}$ is the energy in the largest space of
the ladder ($K_{\max}=60$, $52\,948$ states),
$\Delta_{\rm geo}$ the geometric residual summed from the last
increments, and $E_{\rm ref}=E_{\rm last}-\Delta_{\rm geo}$ the
reference energy used below.  $\delta B$ is the difference between the corresponding
binding energy $B_{\rm ref}=-E_{\rm ref}$ and the published HH value
$B_{\rm lit}$~\cite{Marcucci2020}, quoted there with a $1$-keV uncertainty.}
\squeezetable
\begin{ruledtabular}
\begin{tabular}{lccccc}
Case & $E_{\rm last}$ & $\Delta_{\rm geo}$ &
$E_{\rm ref}$ & $B_{\rm lit}$ & $\delta B$ \\
 & (MeV) & (keV) & (MeV) & (MeV) & (keV) \\
\hline
$^3$H, AV18      & $-7.62138$ & 1.74 & $-7.62312$ & 7.624 & $-0.9$ \\
$^3$He, AV18     & $-6.92285$ & 2.18 & $-6.92503$ & 6.925 & $ 0.0$ \\
$^3$H, AV18+UIX  & $-8.47836$ & 1.16 & $-8.47952$ & 8.479 & $+0.5$ \\
$^3$He, AV18+UIX & $-7.74832$ & 1.43 & $-7.74975$ & 7.750 & $-0.3$ \\
\end{tabular}
\end{ruledtabular}
\end{table}

\begin{figure*}[t]
\centering
\includegraphics[width=0.75\linewidth]{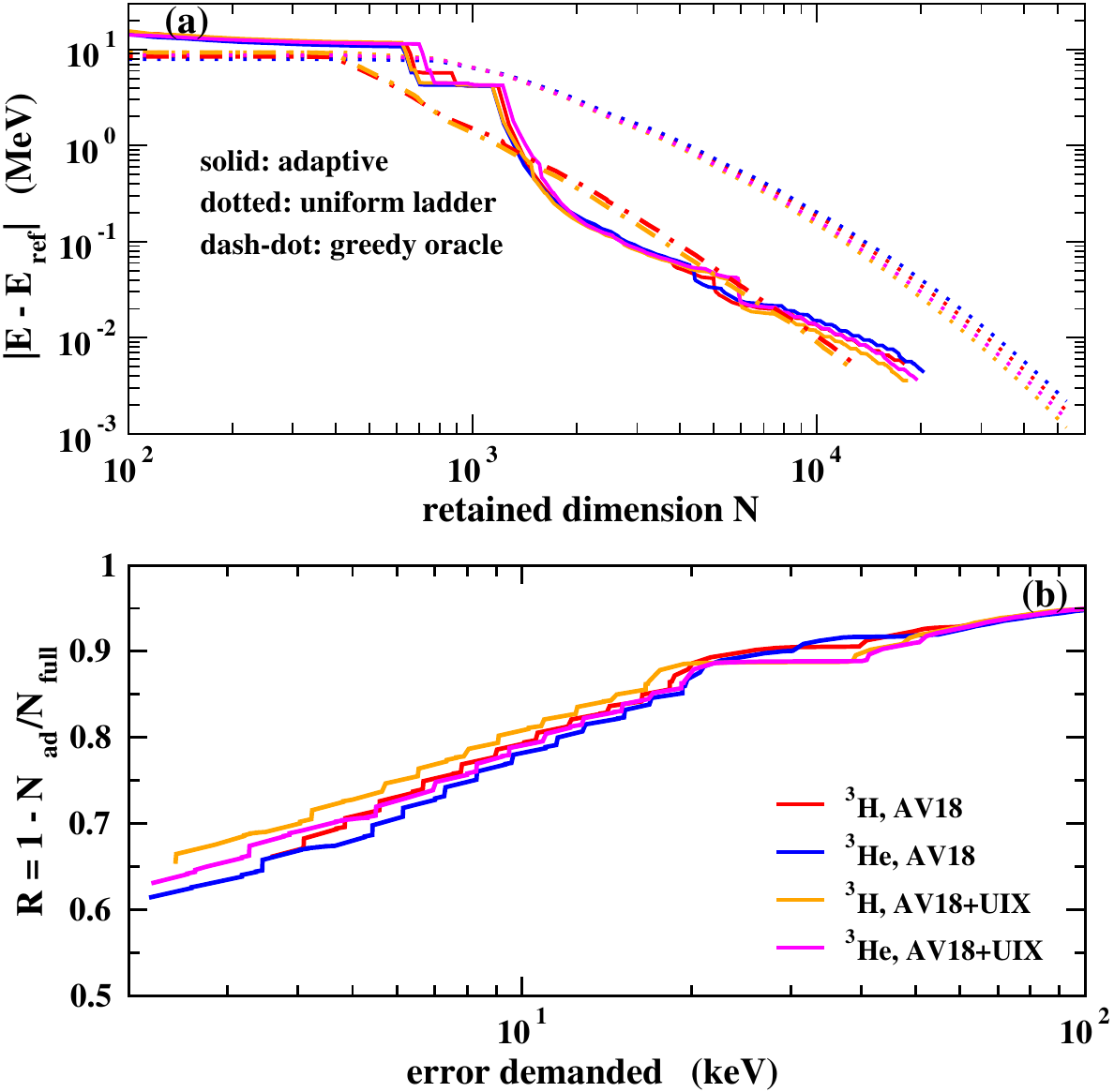}
\caption{\label{fig:pareto}
(a) Energy error relative to $E_{\rm ref}$ against the retained basis
dimension $N$, for the uniform ladder (dotted lines) and the adaptive
construction (solid lines); for $^3$H the dash-dotted lines are the
greedy oracle of Sec.~\ref{sec:results-oracle}.  The last point of the ladder
is the uniform $K=60$ space, $N_{\rm full}=52\,948$.  (b) The reduction
$R=1-N_{\rm ad}/N_{\rm full}$ of the adaptive runs against the accuracy
$\epsilon$ demanded, with the error measured from the uniform $K=60$ space as
in Table~\ref{tab:matched-accuracy}.  Colors label the four
nucleus--Hamiltonian combinations in both panels, as in
Fig.~\ref{fig:reference-convergence}.}
\end{figure*}

\subsection{\label{sec:results-adaptive}Compression of the HH space}

We now compare the ladder with the baseline adaptive runs, i.e.\ the
class-radial calculation for $^3$H and the cold-start calculation for $^3$He,
both free to grow every class to $K_{\max}$.  Fig.~\ref{fig:pareto}(a) shows
the absolute energy error relative to $E_{\rm ref}$ against the retained
dimension, while Tables~\ref{tab:matched-accuracy}
and~\ref{tab:adaptive-summary} collect the numbers.

All four nucleus--Hamiltonian runs satisfy both families of tests of
Sec.~\ref{sec:stopping} and are therefore certified, see
Table~\ref{tab:adaptive-summary}.  What a run certifies is its distance from
the uniform $K_{\max}$ space, the largest space it may reach.  That distance
can be measured here because the uniform ladder was run with the same
Hamiltonian and numerics: the difference between the adaptive energy and
that of the uniform $K_{\max}$ space, $\delta E_{K_{\max}}=E_{\rm ad}-E_{\rm
last}$, isolates the error of the selection alone.  It goes from $2.2$~keV ($^3$He, AV18) to $3.6$~keV ($^3$H, AV18), inside the requested $\epsilon_E=5$~keV in all four cases.
The error relative to the extrapolated reference of
Table~\ref{tab:reference-energies}, $\delta E=\delta E_{K_{\max}}+\Delta_{\rm
geo}$, goes from $3.6$~keV ($^3$H, AV18+UIX) to $5.3$~keV ($^3$H, AV18): it also contains the residual
convergence of the ladder beyond $K_{\max}$, which no space inside $K_{\max}$
can recover.

The final spaces hold on average $19\,000$ states, a little over a third of
the $52\,948$ of the uniform $K=60$ space: the four baseline runs, $^3$H and
$^3$He with AV18 and with AV18+UIX, reproduce that
space within the requested accuracy, and certify that they do, leaving
unbuilt from $61$\% ($^3$He, AV18) to $66$\% ($^3$H, AV18) of it.  Since the Hamiltonian matrix grows with
the square of the dimension, the selected spaces hold about an eighth of the
matrix elements of the $K_{\max}$ space.  In both pairs the $^3$He space is
larger than the $^3$H one, by $14$\% with AV18 and $7$\% with AV18+UIX.  The
final vectors show where the difference comes from: in $^3$He the learner
takes the $S$-wave class two to four shells further, to $K_c=60$ (AV18) and
$58$ (AV18+UIX) against $56$ in $^3$H, and the $L=0$, $T=3/2$ class to
$K_c=18$ and $20$ against $8$ and $12$, with more radial functions.  That
class is populated by isospin breaking, three times more in $^3$He, where the
Coulomb repulsion between the two protons adds to the charge dependence of
the strong interaction, as discussed in Sec.~\ref{sec:results-importance}.

The reduction depends strongly on how much accuracy is demanded, as shown in
Table~\ref{tab:matched-accuracy} and Fig.~\ref{fig:pareto}(b).  Where
$100$~keV is enough, the adaptive space is $95$\% smaller than the $K_{\max}$
space in all four cases; at $20$~keV it is from $87$\% ($^3$He, both
Hamiltonians) to $89$\% ($^3$H, AV18+UIX) smaller, and at $5$~keV from
$67$\% ($^3$He, AV18) to $73$\% ($^3$H, AV18+UIX).  These values refer to the first space of each run that
comes within the given accuracy of the $K_{\max}$ space.  The run cannot know
when it reaches that space, since it does not know $E_{\rm last}$: it goes on
until its own tests certify the accuracy, and stops at the somewhat larger
spaces of Table~\ref{tab:adaptive-summary}, from $61$\% to $66$\% smaller than
the $K_{\max}$ space.  Selection pays when a calculation may stop well short of
exhausting the space, since a few dominant directions then carry almost all
the binding and the rest can be left coarse.  As the tolerance tightens, more
directions must be developed, the adaptive space approaches the uniform one,
and the advantage decreases.  Furthermore, the reduction $R'$ against the
uniform space that reaches the same accuracy shows that the adaptive space is
$80$\% smaller at $100$~keV ($78$\% for $^3$H with AV18+UIX), from $72$\%
(AV18+UIX, both nuclei) to $76$\% ($^3$He, AV18) smaller at $20$~keV, and
from $54$\% ($^3$He, AV18+UIX) to $61$\% ($^3$H, AV18) smaller at $5$~keV.  The two Hamiltonians compress alike.

With the AV18+UIX interaction the $P$-wave class $(1,1,1)$ is worth
$387$~keV, see Sec.~\ref{sec:results-importance}, but the first time the
learner reads it the state is not yet bound, and in that reading the class
gains only $0.05$~keV.  That reading is retired when the state binds; the
class is read again at some $900$ states, where it gains $109$~keV, and it is
developed from there.  If the early reading had been kept, the learner,
which ranks the moves by gain per unit cost, would have left the class aside
until the $S$ and $D$ waves were fully developed.

\begin{table*}[t]
\caption{\label{tab:matched-accuracy}
Basis reduction against the accuracy demanded.  For the four
nucleus--Hamiltonian combinations and for each accuracy $\epsilon$, we list
the dimension of the HH space needed to reach the energy of the uniform
$K=60$ space, of $N_{\rm full}=52\,948$ states, within $\epsilon$, in two
cases: the first space of the uniform ladder ($N_{\rm unif}$) and the
smallest space of the adaptive run ($N_{\rm ad}$).  We also give two
reductions: $R=1-N_{\rm ad}/N_{\rm full}$, against the uniform $K=60$ space,
and $R'=1-N_{\rm ad}/N_{\rm unif}$, against the uniform space of the same
accuracy.  The adaptive entries are read along the single run made with
$\epsilon_E=5$~keV, whose certified final spaces are given in
Table~\ref{tab:adaptive-summary}.}
\squeezetable
\begin{ruledtabular}
\begin{tabular}{l@{\hspace{8pt}}cccc@{\hspace{14pt}}cccc@{\hspace{14pt}}cccc@{\hspace{14pt}}cccc}
 & \multicolumn{4}{c}{$^3$H, AV18} & \multicolumn{4}{c}{$^3$He, AV18}
 & \multicolumn{4}{c}{$^3$H, AV18+UIX} & \multicolumn{4}{c}{$^3$He, AV18+UIX} \\
$\epsilon$ & $N_{\rm unif}$ & $N_{\rm ad}$ & $R$ & $R'$ & $N_{\rm unif}$ & $N_{\rm ad}$ & $R$ & $R'$ &
 $N_{\rm unif}$ & $N_{\rm ad}$ & $R$ & $R'$ & $N_{\rm unif}$ & $N_{\rm ad}$ & $R$ & $R'$ \\
\hline
$100$~keV & 13888 & 2728 & 95\% & 80\% & 13888 & 2812 & 95\% & 80\% & 12180 & 2708 & 95\% & 78\% & 13888 & 2710 & 95\% & 80\% \\
$ 50$~keV & 17640 & 4064 & 92\% & 77\% & 19684 & 4316 & 92\% & 78\% & 15708 & 4464 & 92\% & 72\% & 17640 & 4798 & 91\% & 73\% \\
$ 20$~keV & 24108 & 6398 & 88\% & 73\% & 28980 & 7010 & 87\% & 76\% & 21840 & 6068 & 89\% & 72\% & 24108 & 6780 & 87\% & 72\% \\
$ 10$~keV & 34300 & 11338 & 79\% & 67\% & 34300 & 11658 & 78\% & 66\% & 28980 & 10484 & 80\% & 64\% & 28980 & 11242 & 79\% & 61\% \\
$  5$~keV & 40068 & 15564 & 71\% & 61\% & 40068 & 17264 & 67\% & 57\% & 34300 & 14428 & 73\% & 58\% & 34300 & 15610 & 71\% & 54\% \\
\end{tabular}
\end{ruledtabular}
\end{table*}

\begin{table*}[t]
\caption{\label{tab:adaptive-summary}
Baseline adaptive results, i.e.\ the results obtained when the runs are
asked for $\epsilon_E=5$~keV.  We list the adaptive energy $E_{\rm ad}$; its
distance $\delta E_{K_{\max}}=E_{\rm ad}-E_{\rm last}$ from the energy
$E_{\rm last}$ of the uniform $K_{\max}$ space; the certified residual
$\Delta_{\rm cert}$, the estimate of that distance made by the run itself, which has to
lie below $\epsilon_E$ for the run to stop; the retained dimension
$N_{\rm ad}$; its reduction $R$ with respect to the uniform $K=60$ space of
$52\,948$ states; and the number of exact solves $N_{\rm solve}$.  All four
runs are certified.}
\begin{ruledtabular}
\begin{tabular}{lccccccc}
System & Hamiltonian & $E_{\rm ad}$ (MeV) &
$\delta E_{K_{\max}}$ (keV) & $\Delta_{\rm cert}$ (keV) & $N_{\rm ad}$ & $R$ &
$N_{\rm solve}$ \\
\hline
$^3$H  & AV18      & $-7.61780$ & $+3.58$ & 4.98 & $17\,946$ & 66.1\% & 156 \\
$^3$He & AV18      & $-6.92068$ & $+2.18$ & 4.06 & $20\,452$ & 61.4\% & 170 \\
$^3$H  & AV18+UIX  & $-8.47594$ & $+2.42$ & 3.87 & $18\,370$ & 65.3\% & 163 \\
$^3$He & AV18+UIX  & $-7.74613$ & $+2.20$ & 4.85 & $19\,570$ & 63.0\% & 171 \\
\end{tabular}
\end{ruledtabular}
\end{table*}

\subsubsection{Is the adaptive radial direction worth it?}

Learning the radial resolution of each class, besides its angular cutoff,
reduces the space further, at the price of more exact solves.  To measure
both effects, a companion run on $^3$H with AV18 uses the uniform-radial
mode, in which every sector keeps all $28$ Laguerre functions and only the
eight hyperangular cutoffs are learned; everything else is identical to the
class-radial baseline.  The two runs end certified.  The class-radial run ends
$3.58$~keV from the $K_{\max}$ space with $17\,946$ states, the
uniform-radial one $1.26$~keV from it with $22\,736$ states: the
class-radial space is $21$\% smaller and $2.3$~keV less accurate, both within
the tolerance.  At equal accuracy the gain is largest where the tolerance is
loose.  The reduction $R$ of the class-radial and of the uniform-radial run
is $95$\% and $85$\% at $100$~keV, $92$\% and $82$\% at $50$~keV, $88$\% and
$78$\% at $20$~keV, $79$\% and $71$\% at $10$~keV, and $71$\% and $66$\% at
$5$~keV.  At a loose tolerance most classes need only a few Laguerre
functions, which a single $N_\rho$ cannot provide.  The final radial profile
of the class-radial run shows the same: the $S$ wave keeps the full
$N_\rho=28$ and the $D$ wave $26$, the two $P$ waves $22$ and $24$, and the
four $T=3/2$ classes, below $10^{-4}$ of the norm, $16$ for $(0,1,3)$ and $18$
for the other three.
The price of the eight radial coordinates is the number of exact solves,
$156$ against $107$.  Therefore, we can conclude that indeed the adaptive
radial direction is worth it.  For instance, for $49$ more exact solves, learning the radial resolution of each class
gives a final space $21$\% smaller and, at equal accuracy, ten points more of
reduction at $100$~keV and five at $5$~keV.

We measure the cost of a run mainly by its number of exact solves, the
quantity that further developments of the learner, such as a prior learned
with a neural network as discussed in Sec.~\ref{sec:conclusions}, aim to reduce.  With only eight
classes, $A=3$ is a small selection problem: the learner has few directions
to rank, and the uniform ladder is still affordable.  We count only the
solves of rank-increasing refinements, since a rank-neutral refinement needs
no diagonalization.  The uniform ladder reaches $K_{\max}$ in $26$ solves,
while the four adaptive runs take from $156$ ($^3$H, AV18) to $171$ ($^3$He,
AV18+UIX), as listed in Table~\ref{tab:adaptive-summary}.  Most of the cost
comes from the largest spaces, where the cost of a solve grows faster than
quadratically with the dimension: above some $6000$ states the ladder makes
$16$ solves and the adaptive runs from $54$ ($^3$H, AV18) to $73$ ($^3$He,
AV18+UIX).  Every adaptive move adds a single shell, one
class by $\Delta K=2$ or one radial coordinate by $\Delta N_\rho=2$ Laguerre
functions.  In every run the number of solves equals the number of
shells that separate the starting vector from the final one: $156$,
$170$, $163$ and $171$ for the four runs of Table~\ref{tab:adaptive-summary},
and $107$ for the uniform-radial variant.  The learner therefore never
entered a direction that it later abandoned.

Solves can be saved by moves that add several shells at once, while the gain is
still far above the tolerance; this coarsens the path but leaves every
accepted space exact.  On a reduced $^3$H problem (AV18, $T=1/2$ only,
uniform ladder up to $K=24$, $20$~keV tolerance), single-shell moves take
$69$ exact solves.  Moves of up to six shells, taken only while the
lower bound of the surrogate on the next shell exceeds five times the tolerance,
take $62$, and moves of up to eight, with a threshold of twice the
tolerance, take $47$, a third fewer.  All of them reach the same final space,
of $3862$ states, with the same energy, certified.  What a wide move gives up
are the spaces between its ends, which are never solved: the path is coarser,
and the smallest space reaching a given looser accuracy is known less
precisely.  The production runs make single-shell moves, which is what
resolves the curves of Fig.~\ref{fig:pareto} and the entries of
Table~\ref{tab:matched-accuracy}; a calculation that needs only the final
space can afford a wider step.

\subsection{\label{sec:results-oracle}Distance from the best reachable space}

How much smaller the selected space could have been is measured, for $^3$H,
by the greedy oracle described in Sec.~\ref{sec:setup-diagnostics}.
Table~\ref{tab:oracle} compares the learner and the greedy oracle at equal
accuracy, measured as in Table~\ref{tab:matched-accuracy}, and the greedy oracle
path is the dash-dotted curve of Fig.~\ref{fig:pareto}(a).  The two curves
cross.  Down to $20$~keV the learner reaches each accuracy with the smaller
space: at $100$~keV its space is $31$\% smaller than that of the greedy oracle with AV18
and $28$\% smaller with AV18+UIX, which is possible because the greedy oracle is
greedy.  From $10$~keV on, the space of the learner is the larger one, by $14$\%
(AV18) and $11$\% (AV18+UIX) at $10$~keV, and by $30$\% and $22$\% at
$5$~keV.  The two stop at similar distances from the $K_{\max}$ space: the
greedy oracle at $3.3$~keV (AV18) and $3.4$~keV (AV18+UIX), with $13\,040$ and
$12\,812$ states, and the learner at $3.6$ and $2.4$~keV, with $17\,946$ and
$18\,370$ states.  The final space of the learner is therefore about $2/5$
larger, because the learner goes on beyond the space at which $5$~keV is
first reached, until it can certify it.  The greedy oracle paid for its knowledge
with $1233$ (AV18) and $1236$ (AV18+UIX) exact solves, eight times the $156$
and $163$ of the learner.

\begin{table}[t]
\caption{\label{tab:oracle}
Learner against the greedy oracle, for $^3$H with the AV18 and AV18+UIX
interactions.  For each accuracy $\epsilon$, measured from the uniform $K=60$
space, we list the dimension $N_{\rm ad}$ of the smallest space of the
adaptive trajectory reaching it, the same as in Table~\ref{tab:matched-accuracy},
the dimension $N_{\rm or}$ of the smallest space of the greedy oracle path reaching
it, and their ratio $N_{\rm ad}/N_{\rm or}$.  Note that a ratio below one
means that the adaptive space is the smaller.}
\squeezetable
\begin{ruledtabular}
\begin{tabular}{l@{\hspace{8pt}}ccc@{\hspace{14pt}}ccc}
 & \multicolumn{3}{c}{AV18} & \multicolumn{3}{c}{AV18+UIX} \\
$\epsilon$ & $N_{\rm ad}$ & $N_{\rm or}$ & ratio & $N_{\rm ad}$ & $N_{\rm or}$ & ratio \\
\hline
$100$~keV & 2728  & 3928  & 0.69 & 2708  & 3778  & 0.72 \\
$ 50$~keV & 4064  & 5314  & 0.76 & 4464  & 4780  & 0.93 \\
$ 20$~keV & 6398  & 7240  & 0.88 & 6068  & 7126  & 0.85 \\
$ 10$~keV & 11338 & 9946  & 1.14 & 10484 & 9452  & 1.11 \\
$  5$~keV & 15564 & 11978 & 1.30 & 14428 & 11830 & 1.22 \\
\end{tabular}
\end{ruledtabular}
\end{table}

The final vectors show where the difference between the two spaces lies.
On the angular side they nearly coincide.  The greedy oracle stops the $S$-wave
class at $K=54$ and the $D$ wave at $50$ (AV18) and $48$ (AV18+UIX), where the
learner reaches $56$ and $48$ with both interactions; it stops the $P$ waves
at $26$, the learner at $22$ (AV18) and $24$ (AV18+UIX).  These angular
differences add up to $-12$ states, out of a total difference of $4906$
states with AV18, and to $440$ out of $5558$ with AV18+UIX.  The difference is
radial: the greedy oracle keeps $8$ Laguerre functions on the $P$ waves ($10$ on
$(1,1,1)$ with AV18+UIX), $6$ on the four $T=3/2$ classes and $14$ on the $D$
wave, where the learner keeps $22$ and $24$ on the two $P$ waves with AV18,
$22$ and $26$ with AV18+UIX, $16$ (AV18) or $20$ (AV18+UIX) on the $L=0$,
$T=3/2$ class, $18$ on the other three $T=3/2$ classes and $26$ on the $D$
wave; only on the $S$
wave do both use the full $28$.  The $D$ wave alone accounts for $3600$
states, about three quarters of the difference with AV18 and two thirds with
AV18+UIX.  The reason is the first measured test of Sec.~\ref{sec:stopping}.
Before leaving a coordinate coarse, the learner must measure, three times in
a row, that growing it adds less than $\epsilon_E$ to the binding, and each of
these measurements requires actually growing the coordinate by one shell.
The greedy oracle knows the gains of all moves in advance and needs no such
measurements.  The excess is therefore the resolution the learner spends to establish,
without knowing the answer, that the coordinates it leaves coarse can indeed
be left coarse.  A stopping rule that could establish this more cheaply, such
as the second-order estimate discussed in Sec.~\ref{sec:conclusions}, would
recover part of it.

The rule without a surrogate described in Sec.~\ref{sec:setup-diagnostics}
ranks each coordinate by the gain it gave the last time it was grown,
divided by the cost of growing it again.  It knows less than the learner, and
gives the opposite comparison.  While the gains are large the two follow the same
path: down to $10$~keV the smallest space of each accuracy is the same, see
Table~\ref{tab:matched-accuracy}, because the last measured gain ranks the
moves as the surrogate does.  They part where the gains approach the
tolerance and the run has to establish that its coordinates are spent.  The
rule certifies with $20\,852$ states after $173$ exact solves, $71$ of them
above $6000$ states, in $12.3$ minutes of solve time; the learner with
$17\,946$ states after $156$ solves, $54$ of them above $6000$ states, in
$7.4$ minutes.  The rule takes every $T=1/2$ class one or two shells further
and the $L=0$, $T=3/2$ class four, carries two to eight more Laguerre
functions on four classes, and ends $2.0$~keV from the $K_{\max}$ space,
while the learner ends at $3.6$~keV.  The surrogate thus earns a certified
space $14$\% smaller, a tenth fewer exact solves and two fifths less solve
time, in the part of the run that costs most.

\subsection{\label{sec:results-uq}The residual-basis estimate}

In a predictive application the converged energy is not known.  Here the
uniform ladder provides it, so the stopping tests can be checked against it.
The certified residual $\Delta_{\rm cert}$ of Eq.~(\ref{eq:certified-bound})
estimates what the $K_{\max}$ space still holds beyond the selected one, so
it has to be at least as large as $\delta E_{K_{\max}}$.  It is in all four
nucleus--Hamiltonian runs, see Table~\ref{tab:adaptive-summary}, exceeding the
actual difference by a factor from $1.4$ ($^3$H, AV18) to $2.2$ ($^3$He,
AV18+UIX), while using three
quarters or more of the $5$~keV budget.  The certificate does not cover the
tail beyond $K_{\max}$, $\Delta_{\rm geo}$, discussed in
Sec.~\ref{sec:results-adaptive}.  A calculation that needs the accuracy
$\epsilon_E$ with respect to the converged energy, rather than to the
$K_{\max}$ space, must estimate that tail separately, from a uniform ladder
as done here, or use a larger $K_{\max}$.

The certified residual is the tail estimate multiplied by the calibration
scale $s_{\rm conf}$ of Eq.~(\ref{eq:certified-bound}), which the run obtains
by comparing its own earlier predictions with the binding gained afterwards,
and which can widen the estimate but never narrow it.  In the four baseline
runs the
calibration never inflated the estimate, i.e.\ $s_{\rm conf}=1$: the estimate
was already cautious, about twice the tail actually observed afterwards.  The
predictions made along each run, from $16$ ($^3$H, AV18) to $27$ ($^3$He,
AV18+UIX), all cover the value
observed afterwards, against a nominal level $\gamma=0.90$.

The settings of the estimate---$\gamma$, the three estimators and the rule
combining them, with the values of Sec.~\ref{sec:setup-protocol}---were
fixed on $^3$H with AV18 and used unchanged in the other three cases: they
carry over to the mirror nucleus and to the Hamiltonian with the
three-nucleon force, and their transfer to other mass numbers remains to be
tested.  The same estimate extended beyond $K_{\max}$ to an infinite basis
gives from $5.3$~keV ($^3$H, AV18+UIX) to $8.2$~keV ($^3$H, AV18).  It is what larger cutoffs could still add, reported
as information: the certification does not depend on it.

To stop, a run must have measured that every coordinate gains less than the
tolerance.  A measurement ages: a hyperradial reading taken before the
hyperangular cutoff of its class grew no longer describes the class as it
is, and the rule of Sec.~\ref{sec:stopping} retires it and has the
coordinate measured again.  A reduced $^3$H problem with AV18 and a
$20$-keV tolerance shows that the rule is needed: without it the run stops at
$143$~keV from the answer, while with it it stops at $16$~keV.

One may ask whether a reading should also be retired when a different class
grows.  The four nucleus--Hamiltonian runs say it should not.  For the hyperangular moves made
on a bound state, the ratio of the measured to the expected gain has median
$0.84$ when neither the $S$- nor the $D$-wave class has grown since the
class was last read, and $0.82$ when one of them has.  Only the frequency of
very small gains changes: the gain falls below a third of the expectation
in $8$\% of the moves in the first case and in $20$\% in the second, the
dominant classes, once developed, leaving less for the others to take.  An
old reading is therefore mostly too high.  The only consequence is that the
run makes a few more solves than needed; the accuracy is not affected.  No rule across classes is applied.

Along a run the one-step predictions improve: the coverage of the calibrated
predictive intervals rises, against a nominal $0.90$, from between $0.29$
($^3$H, AV18) and $0.38$ ($^3$He, AV18+UIX) in the first third of a trajectory
to between $0.59$ ($^3$H, AV18) and $0.77$ ($^3$He, AV18) in the second and
between $0.89$ ($^3$H, AV18) and $0.93$ ($^3$He, AV18+UIX) in the last.  The first third is where the state binds and
every earlier reading is retired; past it the median discrepancy between
predicted and measured gain is at most a factor $2$.  As expected, the
predictions are least reliable early in a run, where they guide the
exploration, and reliable at its end, where they enter the stopping
decision.

\begin{figure}[t]
\centering
\includegraphics[width=0.95\linewidth]{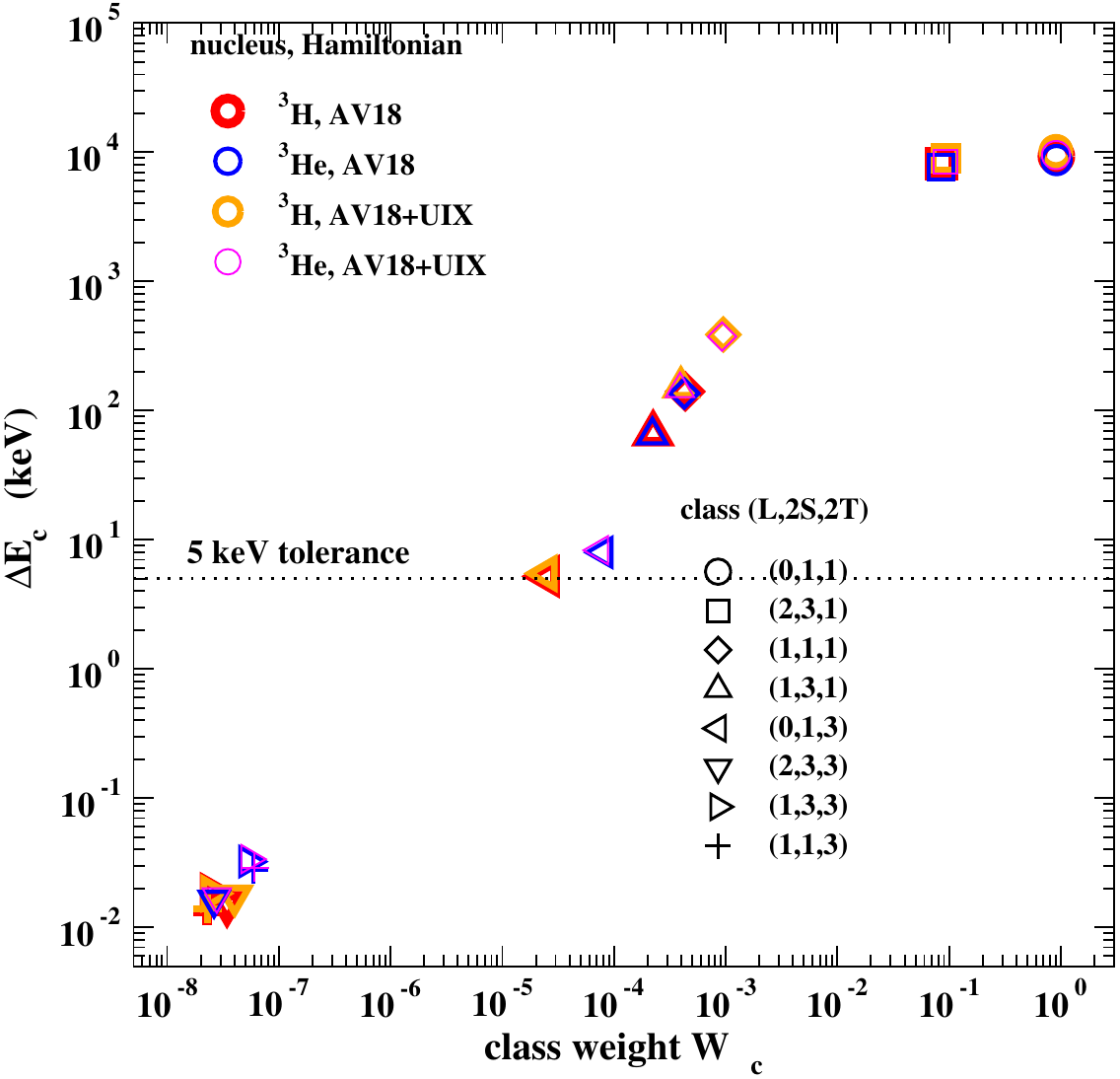}
\caption{\label{fig:weight-importance}
Energy lost on removing a class, $\Delta E_c$, against the probability $W_c$
it carries, for all eight HH classes of each of the four
nucleus--Hamiltonian combinations.  Color identifies the combination, as in
Fig.~\ref{fig:reference-convergence}, and symbol shape the class
$(L,2S,2T)$; the four combinations agree closely enough that each class forms
a single cluster, and the markers are drawn hollow, with a stroke of
different width for each combination, so that those lying on top of one another remain visible.  The
relation spans eight decades in weight and six in energy and is far from
proportional: $(1,1,1)$ and $(1,3,1)$, near $W_c\sim10^{-4}$, are worth
$10^2$~keV, and $(0,1,3)$ sits at the $5$~keV tolerance (dashed line).  Only
the three classes below $10^{-7}$ fall clearly beneath it, so a truncation on
probability alone would have to set its threshold where weight no longer
predicts importance.  For the two dominant classes, at the top right, the
reduced space has no bound state, so their ordinate is not a binding loss
and is set in part by the hyperradial basis; the argument here concerns the
classes below them.}
\end{figure}

\subsection{\label{sec:results-importance}Class composition and the effect of
the three-nucleon interaction}
\label{sec:results-uix}

In Fig.~\ref{fig:weight-importance} we show $\Delta E_c$ against $W_c$ for the
final reference states, and Table~\ref{tab:class-composition} lists both.
The composition is the expected one.  For $^3$H with AV18 the $S$-wave class
$(0,1,1)$ carries $W_c=0.914$ and the tensor-induced $D$-wave class $(2,3,1)$
$0.0851$, i.e.\ the familiar $8.5$\% $D$-state probability; the two $P$-wave
classes together carry $6.6\times10^{-4}$, and the $T=3/2$ admixtures
generated by charge dependence $2.5\times10^{-5}$ in the $L=0$ class and less
than $10^{-7}$ in the rest.  $^3$He is its mirror to three digits, except
that the Coulomb interaction raises the $L=0$, $T=3/2$ admixture to
$8.0\times10^{-5}$.  Table~\ref{tab:probabilities} compares these
probabilities with those of the calculations of Ref.~\cite{Nogga2003}.  The $P$-wave and $T=3/2$ probabilities agree to the last digit quoted there
and the $D$-wave ones within $0.002$.  The probability of the $S'$ component, the $S$ wave of mixed spatial
symmetry, is obtained by projecting the $L=0$, $T=1/2$ class of the $K_{\max}=60$
state on the permutation symmetry of its spatial part; it is lower than
that of Ref.~\cite{Nogga2003} by $0.009$ ($^3$H, AV18+UIX) to $0.027$
($^3$He, AV18), in percent, while the sum of
the two $S$-wave components agrees.  The $S'$ component is the slowest to
converge of the four, still growing at $K=60$, so the value quoted is that
of the $K_{\max}$ space and not an extrapolation.

By inspection of Fig.~\ref{fig:weight-importance}, we can see that energetic importance
follows a different order.  Removing either the $S$- or
the $D$-wave class leaves no bound state at all: a class carrying a tenth of
the norm is as indispensable as the one carrying nine tenths, as expected
from the role of the tensor force.  Below them the $P$-wave class $(1,1,1)$
carries $4.3\times10^{-4}$ of the norm, two orders of magnitude less than
the $D$ wave, yet removing it costs $140$~keV; the $L=0$, $T=3/2$ class
carries $2.5\times10^{-5}$ and is still worth $5.2$~keV, exactly at the
tolerance; only the three $L\geq1$, $T=3/2$ classes, at $10^{-8}$ of the norm
and with removal energies of $0.01$~keV for $(1,1,3)$ and $0.02$~keV for the
other two, are negligible.  A truncation on probability would
have to keep every class down to $2.5\times10^{-5}$ of the norm to stay
within $5$~keV, a threshold known only once the removal energies are
computed.  Saturation is a separate question: at the final
reference point every class has $\Delta E_c^{\rm top}<0.5$~keV.  The learner
reproduces this hierarchy from marginal gains alone.  For $^3$H with AV18,
with every class free to reach $K_{\max}=60$, its final vector stops at
$K_c=56$ and $48$ for the $S$- and $D$-wave classes, at $22$ for both $P$
waves, at $8$ for the $L=0$, $T=3/2$ class, and at $14$ for $(1,1,3)$ and
$(1,3,3)$ and $12$ for $(2,3,3)$, the few shells it takes to show that they have
nothing to give: occupation and marginal relevance are distinct.

\begin{table*}[t]
\caption{\label{tab:class-composition}
Class composition of the converged reference states.  $W_c$ is the
probability carried by the class and $\Delta E_c$ the binding lost when the
class is removed and the space rediagonalized; the latter are not additive.
Removing either of the two dominant classes leaves no bound state, so no
$\Delta E_c$ is quoted for them: the lowest eigenvalue of the reduced space is
then a discretized continuum level, fixed by the hyperradial basis rather
than by the Hamiltonian.
Classes are ordered by weight in $^3$H with AV18.}
\begin{ruledtabular}
\begin{tabular}{l@{\hspace{20pt}}cc@{\hspace{30pt}}cc}
 & \multicolumn{2}{c}{$^3$H, AV18} & \multicolumn{2}{c}{$^3$H, AV18+UIX} \\
$(L,2S,2T)$ & $W_c$ & $\Delta E_c$ (keV) & $W_c$ & $\Delta E_c$ (keV) \\
\hline
$(0,1,1)$ & $9.14\times10^{-1}$ & unbound & $9.06\times10^{-1}$ & unbound \\
$(2,3,1)$ & $8.51\times10^{-2}$ & unbound & $9.30\times10^{-2}$ & unbound \\
$(1,1,1)$ & $4.35\times10^{-4}$ & 140.2 & $9.52\times10^{-4}$ & 387.4 \\
$(1,3,1)$ & $2.24\times10^{-4}$ & 66.3 & $3.98\times10^{-4}$ & 151.8 \\
$(0,1,3)$ & $2.51\times10^{-5}$ & 5.2 & $2.48\times10^{-5}$ & 5.5 \\
$(2,3,3)$ & $3.43\times10^{-8}$ & 0.02 & $4.00\times10^{-8}$ & 0.02 \\
$(1,3,3)$ & $2.54\times10^{-8}$ & 0.02 & $2.51\times10^{-8}$ & 0.02 \\
$(1,1,3)$ & $2.28\times10^{-8}$ & 0.01 & $2.29\times10^{-8}$ & 0.01 \\
\end{tabular}
\end{ruledtabular}
\end{table*}

\begin{table}[t]
\caption{\label{tab:probabilities}
Probabilities, in percent, of the $S'$, $P$ and $D$ components and of the
$T=3/2$ admixture in the uniform $K_{\max}=60$ states, against those of
Ref.~\cite{Nogga2003}.}
\squeezetable
\begin{ruledtabular}
\begin{tabular}{lccccc}
Case & & $P_{S'}$ & $P_P$ & $P_D$ & $P_{T=3/2}$ \\
\hline
$^3$H, AV18      & this work & 1.278 & 0.066 & 8.512 & 0.0025 \\
                 & Ref.~\cite{Nogga2003} & 1.293 & 0.066 & 8.510 & 0.0025 \\
$^3$H, AV18+UIX  & this work & 1.045 & 0.135 & 9.302 & 0.0025 \\
                 & Ref.~\cite{Nogga2003} & 1.054 & 0.135 & 9.301 & 0.0025 \\
$^3$He, AV18     & this work & 1.503 & 0.065 & 8.469 & 0.0080 \\
                 & Ref.~\cite{Nogga2003} & 1.530 & 0.065 & 8.467 & 0.0080 \\
$^3$He, AV18+UIX & this work & 1.225 & 0.132 & 9.250 & 0.0075 \\
                 & Ref.~\cite{Nogga2003} & 1.242 & 0.132 & 9.248 & 0.0075 \\
\end{tabular}
\end{ruledtabular}
\end{table}

Because the AV18 and AV18+UIX references use identical spaces, the effect of
the three-nucleon interaction can be examined without changing
representation.  At the final space
$\Delta E_{\rm UIX}(\mathcal{V})=E_{\rm AV18+UIX}(\mathcal{V})-E_{\rm AV18}(\mathcal{V})$
is $-857.0$~keV in $^3$H and $-825.5$~keV in $^3$He, while the expectation
values of the two UIX components on the converged state,
$\langle V_{2\pi}\rangle=-2129.8$ and $\langle V_R\rangle=+1001.7$~keV in
$^3$H ($-2045.4$ and $+958.0$~keV in $^3$He), sum to $-1128.0$~keV for $^3$H
and $-1087.4$~keV for $^3$He.  The differences, $271$~keV in $^3$H and
$262$~keV in $^3$He, are the relaxation of the wave function, and a space selected for AV18 has to be revalidated when
UIX is added.  The Green's function
Monte Carlo values for $^3$H with the same Hamiltonian,
$\langle V_{2\pi}\rangle=-2.2(1)$~MeV for the two-pion-exchange component and
$-1.20(7)$~MeV for the full three-nucleon interaction,
$\langle V_{\rm UIX}\rangle=\langle V_{2\pi}\rangle+\langle V_R\rangle$~\cite{Pudliner1997},
are compatible with ours, $-2.13$ and $-1.13$~MeV respectively, at the level
of their statistical errors.  UIX
also moves importance outwards.  On the common space
the tensor $D$-wave weight rises from $0.0851$ to $0.0930$ at the expense of
the $S$ wave, and the removal energy of the $P$-wave class $(1,1,1)$ grows
from $140$ to $387$~keV, a factor of $2.8$, with $(1,3,1)$ following at
$2.3$, while the $T=3/2$ admixtures are untouched to the digit.  The learner
follows: with UIX the two $P$-wave classes are taken to $K_c=24$ instead of
$22$ and $(1,3,1)$ to $26$ Laguerre functions instead of $24$, while the
$S$- and $D$-wave classes end at the same cutoffs under both Hamiltonians.

\subsection{\label{sec:results-transfer}Transfer between calculations}

\begin{figure}[t]
\centering
\includegraphics[width=0.95\linewidth]{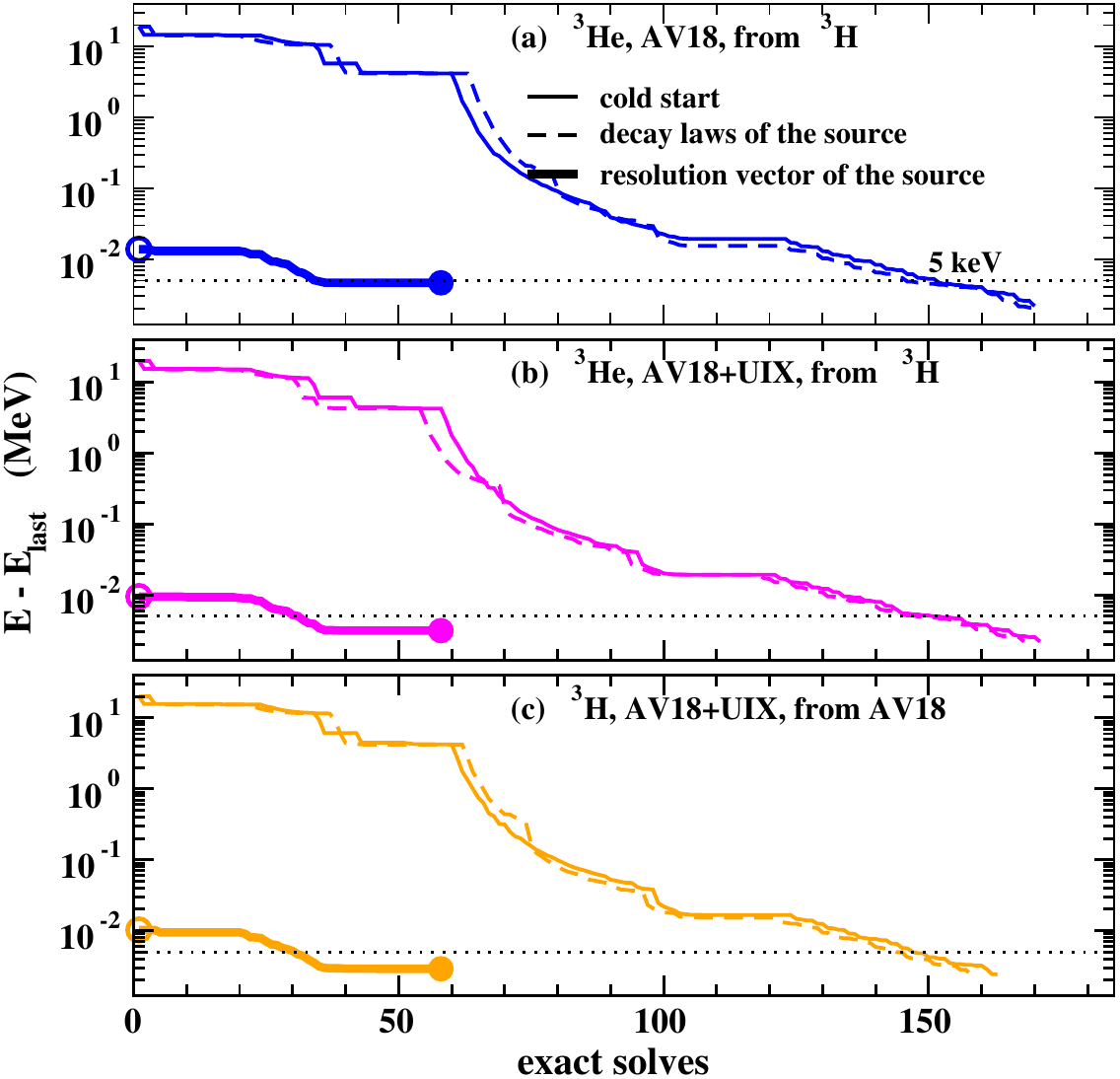}
\caption{\label{fig:transfer}
Distance from the energy of the uniform $K=60$ space against the number of
exact solves made, for the three transfer tests of Table~\ref{tab:transfer}:
(a) $^3$He with AV18, starting from $^3$H with AV18; (b) $^3$He with
AV18+UIX, starting from $^3$H with AV18+UIX; (c) $^3$H with AV18+UIX, starting
from $^3$H with AV18.  Each target is started in three different ways: from
the smallest space (thin solid curve), with the surrogate initialized by the
decay laws of the source (dashed curve), and from the final resolution vector
of the source taken three shells back (thick solid curve, from the open
to the full circle).  The first two are indistinguishable on the scale of
the plot.  The third one starts from the distance reached by the others after
about $130$ solves, and its flat end corresponds to the readings that certify
it.  The dotted line marks the requested accuracy of $\epsilon_E=5$~keV.}
\end{figure}

Table~\ref{tab:transfer} collects the results of the three transfer tests
introduced in Sec.~\ref{sec:setup-diagnostics}: $^3$He from $^3$H, with AV18
and with AV18+UIX, and $^3$H with AV18+UIX from $^3$H with AV18.  Each target
is run from a cold start, with the decay laws of the source, and finally from
the resolution vector of the source.  Fig.~\ref{fig:transfer} follows the
nine runs solve by solve: each panel shows one case, with the distance from
the energy of the $K_{\max}=60$ space after each exact solve, for the three
starts.

By inspection of Fig.~\ref{fig:transfer}, we can see that the run
initialized with the decay laws stays close to the cold start, as
anticipated in Sec.~\ref{sec:transfer-method}.  As shown in
Table~\ref{tab:transfer}, from $^3$H to $^3$He the cold start and the
initialized run both need $170$ exact solves with AV18, and $171$ and
$168$ with AV18+UIX, respectively; for $^3$H with AV18+UIX, initialized from
the run with AV18, they need $163$ and $158$.  The final dimensions of the
two runs differ by at most $5$\%.

By inspection of Fig.~\ref{fig:transfer}, we can also see that the run
started from the resolution vector (thick curve) begins at about $10$~keV
from the $K_{\max}=60$ space, in a space of about $10\,500$ states, a
distance that the other two runs reach only after more than $120$ solves.
It reaches the $5$~keV accuracy after ${\sim}30$ solves and then stays flat.
This means that the last solves gain almost nothing: they are the readings
that the measured tests require before a coordinate can be declared spent.
The three runs certify after $58$ exact solves each, a third of those of the
cold start, in spaces up to $11$\% smaller and with from $23$\% ($^3$H,
AV18+UIX) to $34$\% ($^3$He, AV18+UIX) less solve time, see
Table~\ref{tab:transfer}, since every solve is made near the end of the path.  Each coordinate is moved three or four times, and the final
vector is that of the source or one shell beyond it.  In the third case
the vector learned with AV18 does not carry what UIX adds, and the run
certifies all the same.

We finally remark that mirror nuclei, and one nucleus with two Hamiltonians,
are favorable cases: the classes are the same and their content nearly so.
Between nuclei of different mass number the source lacks some of the classes
of the target, and how to build the starting vector there remains to be
studied.

\begin{table*}[t]
\caption{\label{tab:transfer}
Transfer between calculations.  Each target is run from a cold start
(cold), with the surrogate initialized by the decay laws of the source
(law), and from the final resolution vector of the source taken three
shells back (vector).
$N_{\rm solve}$ is the number of exact solves, $N_0$ the dimension of the
starting space and $N_{\rm ad}$ that of the final one; $\delta E_{K_{\max}}$
and $\Delta_{\rm cert}$ are the same quantities defined in
Table~\ref{tab:adaptive-summary}, both in keV.  Finally, $t$ is the time
spent in exact solves, in minutes on one node.
All nine runs are certified.}
\squeezetable
\begin{ruledtabular}
\begin{tabular}{@{}l@{\hspace{16pt}}c@{\hspace{16pt}}c@{\hspace{16pt}}c@{\hspace{16pt}}c@{\hspace{16pt}}c@{\hspace{16pt}}c@{\hspace{16pt}}c@{}}
Target & Start & $N_{\rm solve}$ & $N_0$ & $N_{\rm ad}$ & $\delta E_{K_{\max}}$ &
$\Delta_{\rm cert}$ & $t$ \\
\hline
$^3$He, AV18 & cold          & 170 & 6        & $20\,452$ & 2.18 & 4.06 & 12.5 \\
from $^3$H   & law           & 170 & 6        & $20\,524$ & 2.07 & 4.81 & 13.3 \\
             & vector        &  58 & $10\,474$ & $18\,278$ & 4.64 & 4.91 & 8.8 \\
\hline
$^3$He, +UIX & cold          & 171 & 6        & $19\,570$ & 2.20 & 4.85 & 17.6 \\
from $^3$H   & law           & 168 & 6        & $19\,344$ & 2.20 & 4.72 & 17.5 \\
             & vector        &  58 & $10\,686$ & $18\,734$ & 3.13 & 4.94 & 11.6 \\
\hline
$^3$H, +UIX  & cold          & 163 & 6        & $18\,370$ & 2.42 & 3.87 & 14.1 \\
from AV18    & law           & 158 & 6        & $17\,526$ & 2.43 & 4.86 & 13.1 \\
             & vector        &  58 & $10\,474$ & $18\,278$ & 2.90 & 3.15 & 10.9 \\
\end{tabular}
\end{ruledtabular}
\end{table*}

\subsection{\label{sec:results-observables}Observables}
\label{sec:results-experiment}

We consider the one-body observables, i.e.\ the matter, point-proton,
point-neutron and charge radii and the magnetic moment.  The charge radius
is obtained from the point-proton one as
$r_{\rm ch}^2=r_p^2+R_p^2+(N/Z)\langle r_n^2\rangle+3/(4m_p^2)$, with
$R_p=0.84075$~fm~\cite{CODATA2022}, $\langle r_n^2\rangle=-0.1161$~fm$^2$~\cite{PDG2020}
and the Darwin--Foldy term $3/(4m_p^2)=0.0332$~fm$^2$.
Table~\ref{tab:observable-convergence} compares the results for these
observables obtained in the final reference space and in the
energy-selected adaptive space.  By inspection of the table, we
can see that in the space selected for the energy, a third of the $K_{\max}$
space, these five observables agree with the reference to about
$2\times10^{-3}$ in relative terms; the absolute differences (in fm for the
radii and in $\mu_N$ for the magnetic moments) are given in
Table~\ref{tab:observable-convergence}.  Targeting the observables explicitly brings every
entry to $4\times10^{-5}$ or better, in relative terms, two orders of magnitude
inside tolerance with both Hamiltonians, AV18 and AV18+UIX, while the energy
comes within $0.04$~keV of that of the $K_{\max}$ space.  The spaces grow
from $17\,946$ ($18\,370$) to $30\,584$ ($30\,632$) states for $^3$H and from
$20\,452$ ($19\,570$) to $31\,872$ ($31\,272$) for $^3$He with AV18
(AV18+UIX), with $62$ ($57$) and $53$ ($51$) additional solves, so that the observables are
converged to $10^{-5}$ leaving unbuilt from $40$\% ($^3$He, AV18) to $42$\%
($^3$H, both Hamiltonians) of the $K_{\max}$ space.

These runs end on the measured tests of Sec.~\ref{sec:stopping}: the last
changes of every radius are below $10^{-5}$~fm, three orders of
magnitude inside tolerance.  The convergence of an observable is
established by these measurements.  Unlike the energy, an observable need
not converge monotonically, so the extrapolated estimate of its tail is an
upper limit and not a sharp one: for the point-proton radius it lies
between $0.006$~fm ($^3$H, AV18+UIX) and $0.027$~fm ($^3$He, AV18), against
measured changes of
$10^{-6}$~fm and a distance from the reference of $10^{-5}$~fm.

\begin{table*}[t]
\caption{\label{tab:observable-convergence}
One-body observables for the four nucleus--Hamiltonian combinations
(column System).  The column Obs.\ gives the observable: the matter,
point-proton, point-neutron and charge radii, $r_m$, $r_p$, $r_n$ and
$r_{\rm ch}$, and the magnetic moment $\mu$.  $O_{\rm ref}$ is its calculated
value in the uniform $K_{\max}=60$ space, and $O_{\rm exp}$ its measured value
(charge radii from Ref.~\cite{Amroun1994}, magnetic moments from
Ref.~\cite{Stone2005}); a dash marks the observables without a measured
value.  $\epsilon_O$ is the tolerance required on the observable in the
observable-targeted runs, chosen as $2\times10^{-3}$ of $O_{\rm ref}$.
$|\Delta O_E|$ is the absolute value of the difference between the value in
the energy-selected adaptive space and $O_{\rm ref}$ (the difference can have
either sign), and $|\Delta O_{\rm obs}|$ the same quantity in the space
selected with the observables as targets, given in units of $10^{-5}$.  Radii
are given in fm and magnetic moments in nuclear magnetons, for $O_{\rm ref}$,
$O_{\rm exp}$, $\epsilon_O$ and $|\Delta O_E|$ alike.  The magnetic moments
are obtained using the one-body current operator only, so their difference from experiment is dominated by the
two-body currents not included here~\cite{Marcucci2005,Marcucci2016}.}
\squeezetable
\begin{ruledtabular}
\begin{tabular}{@{}l@{\hspace{14pt}}c@{\hspace{14pt}}c@{\hspace{14pt}}c@{\hspace{14pt}}c@{\hspace{14pt}}c@{\hspace{14pt}}c@{}}
System & Obs. & $O_{\rm ref}$ & $O_{\rm exp}$ & $\epsilon_O$ & $|\Delta O_E|$ &
$|\Delta O_{\rm obs}|$ \\
 & & & & & & $(10^{-5})$ \\
\hline
$^3$H, AV18  & $r_m$        & $1.7658$ & ---         & 0.0035 & 0.00318 & 2.6 \\
     & $r_p$        & $1.6514$ & ---         & 0.0033 & 0.00153 & 0.5 \\
     & $r_n$        & $1.8204$ & ---         & 0.0036 & 0.00393 & 4.0 \\
     & $r_{\rm ch}$ & $1.7986$ & $1.755(86)$ & 0.0036 & 0.00140 & 0.5 \\
     & $\mu$        & $2.6045$ & $2.97896$   & 0.0052 & 0.00026 & $<0.1$ \\
\hline
$^3$H, +UIX  & $r_m$        & $1.6811$ & ---         & 0.0034 & 0.00198 & 2.1 \\
     & $r_p$        & $1.5804$ & ---         & 0.0032 & 0.00083 & 0.2 \\
     & $r_n$        & $1.7293$ & ---         & 0.0035 & 0.00251 & 3.1 \\
     & $r_{\rm ch}$ & $1.7337$ & $1.755(86)$ & 0.0035 & 0.00075 & 0.2 \\
     & $\mu$        & $2.5952$ & $2.97896$   & 0.0052 & 0.00002 & 0.3 \\
\hline
$^3$He, AV18 & $r_m$        & $1.8047$ & ---         & 0.0036 & 0.00234 & 2.8 \\
     & $r_p$        & $1.8659$ & ---         & 0.0037 & 0.00312 & 4.1 \\
     & $r_n$        & $1.6754$ & ---         & 0.0034 & 0.00060 & 0.2 \\
     & $r_{\rm ch}$ & $2.0405$ & $1.959(30)$ & 0.0041 & 0.00286 & 3.8 \\
     & $\mu$        & $-1.7439$ & $-2.12762$ & 0.0035 & 0.00031 & 0.1 \\
\hline
$^3$He, +UIX & $r_m$        & $1.7138$ & ---         & 0.0034 & 0.00228 & 2.6 \\
     & $r_p$        & $1.7676$ & ---         & 0.0035 & 0.00293 & 3.8 \\
     & $r_n$        & $1.6007$ & ---         & 0.0032 & 0.00084 & 0.2 \\
     & $r_{\rm ch}$ & $1.9510$ & $1.959(30)$ & 0.0039 & 0.00266 & 3.5 \\
     & $\mu$        & $-1.7378$ & $-2.12762$ & 0.0035 & 0.00009 & $<0.1$ \\
\end{tabular}
\end{ruledtabular}
\end{table*}

Finally, the basis error controlled here has to be compared with the error
associated with the predictions of the Hamiltonian itself, which the
comparison with experiment measures.  The
calculated values quoted here are the references, i.e.\ $E_{\rm ref}$ and the
observables of the $K_{\max}$ space.  With AV18 alone the binding energies of
$^3$H and $^3$He are $7.6231$ and $6.9250$~MeV, against the measured $8.4818$
and $7.7180$~MeV~\cite{Wang2021}, the well-known underbinding of a
two-nucleon interaction.  UIX brings them to $8.4795$ and $7.7497$~MeV, within
a few keV of the measured value for $^3$H, to which its parameters are
fitted, and $32$~keV above it for $^3$He.  The truncation error controlled
here, about $5$~keV, is thus two orders of magnitude below the error of the
two-nucleon Hamiltonian and comparable to the residual error of the fitted
three-nucleon one.  The calculated values of the one-body charge radii with
AV18+UIX, $1.7337$~fm for $^3$H and $1.9510$~fm for $^3$He, agree with the
measured $1.755(86)$ and $1.959(30)$~fm~\cite{Amroun1994} within their
uncertainties.  On the other hand, the calculated one-body magnetic moments of
$^3$H and $^3$He, $2.5952$~$\mu_N$ and $-1.7378$~$\mu_N$, differ from the
measured $2.97896$~$\mu_N$ and $-2.12762$~$\mu_N$~\cite{Stone2005} by $13$\%
and $18$\%.  However, since the moment is converged to $10^{-4}$ within the
representation, this difference is not a basis effect: what we see is rather
a deficiency of the transition operator, which lacks the two-body
currents~\cite{Marcucci2005,Marcucci2016}.

\section{\label{sec:conclusions}Discussion and conclusions}

We have used active learning to distribute resolution inside the
hyperspherical-harmonics representation, letting the calculation decide how
finely each physically identifiable class is resolved instead of enlarging
the space uniformly.  The surrogate selects each extension and the
many-body solver evaluates it, so every reported energy is variational in
an explicitly constructed basis; the learning decides which space is solved
and when the run stops.  The surrogate only has to rank the extensions, and it ranks
better as a run proceeds, the coverage
of its predictions rising from one third in the first third of a trajectory
to the nominal value in the last.

Against uniform ladders that reproduce established HH benchmarks within
$1$~keV, the adaptive runs asked for $5$~keV reproduce the uniform
$K_{\max}$ space within $4$~keV in all four nucleus--Hamiltonian
combinations investigated in this work, with little more than a third of its
states.  The reduction grows as the tolerance loosens, to
$95$\% of that space where $100$~keV is enough; resolving the radial basis
per class increases the reduction by ten points at $100$~keV and by five at
$5$~keV.  The
selected representations reproduce the class hierarchy of the trinucleon
without being informed of it at the beginning.  With every class free to reach $K_{\max}=60$, they
spend their resolution on the $S$ and tensor-induced $D$ classes, stop the
$P$ waves at $K_c=22$ with AV18 and $24$ with AV18+UIX, for the binding they
carry, from $66$~keV for $(1,3,1)$ with AV18 to $387$~keV for $(1,1,1)$ with
AV18+UIX,
and take the three
$L\geq1$, $T=3/2$ classes no further than $K_c=14$.  The
one-body radii and the magnetic moments are converged at about the
$2\times10^{-3}$ level (relative error) already in the energy-selected space.

We have also demonstrated that what one run has learned can initialize
another.  Started from the
resolution reached by the mirror nucleus, or by the same nucleus with the
two-nucleon interaction alone, a run certifies a smaller space with a third
of the exact solves and from $23$\% to $34$\% less solve time, reducing to
the readings that certification requires.  The cases tested here share all
their classes.  What such a start is worth between nuclei of different mass,
which do not, is subject of future work.

The control of the accuracy rests on two kinds of evidence, the gains a run
measures and an extrapolated estimate of what remains, which in all four
runs covers the distance from the uniform $K_{\max}$ space by a factor from
$1.4$ ($^3$H, AV18) to $2.2$ ($^3$He, AV18+UIX).  Two comparisons place the learner.  Against a greedy
search that knows the answer, the space of the learner is the smaller of the two
down to $20$~keV and is $2/5$ larger when certified, the excess being the
hyperradial resolution needed to show that the coordinates left coarse can
indeed be left coarse.  Against a rule that ranks the moves on the last
measured gain, without a surrogate, it follows the same path down to
$10$~keV and certifies a space $14$\% smaller with a tenth fewer exact
solves.

Our study allows us to consider naturally two developments in the near
future.  First of all, a perturbative estimate of the energy
held by the excluded space, in the spirit of importance-truncation
methods~\cite{Roth2007,Roth2009}, would give an accuracy control that does
not depend on the trajectory, at one matrix--vector product per check, and
could replace part of the exact solves.  Second, for classes without a
counterpart in lighter nuclei, a prior learned over class descriptors rather than class
labels, with neural networks of the kind now used for the wave function
itself~\cite{Lovato2022,Gnech2024nqs}, can be considered the natural way to
initialize them.

This work is the first step of a program aimed at nuclei of larger mass.
The construction is formulated for arbitrary $A$, and the three-nucleon
systems were chosen because they admit references of benchmark quality,
against which it is possible to measure the efficiency and accuracy of the
method.  Adaptive calculations for $^4$He, $^6$He and
$^6$Li with a chiral interaction are under way.

\begin{acknowledgments}
The calculations were made possible by grants of computing time from the
Italian National Supercomputing Center CINECA, under the agreement
CINECA-INFN, on the Leonardo supercomputer, owned by the EuroHPC Joint
Undertaking and hosted by CINECA (Italy)~\cite{Leonardo}. 
\end{acknowledgments}

\section*{Data availability}
The data that support the findings of this article are available from the
authors upon reasonable request.

\appendix

\section{\label{app:gp}Gaussian-process specification}

This appendix collects the expressions behind the surrogate of
Sec.~\ref{sec:gp-surrogate}, for a generic coordinate $a\in\mathcal{I}$ with
value $x_a$, either $K_c$ or $N_{\rho,c}$, and refinement index
\begin{equation}
 n_a(\bm{x})=\max\left[0,\,\frac{x_a-x_a^{\min}}{2}+1\right],
 \label{eq:app-refinement-index}
\end{equation}
$x_a^{\min}$ being the value at which the coordinate first becomes active.  The GP expressions are the
standard ones~\cite{Rasmussen2006}; the regularization, the calibration and
the bounds described below, and their numerical values, are choices made for
this work, fixed on $^3$H with AV18 and used unchanged in all runs, as
discussed in Sec.~\ref{sec:results-uq}.

The mean parameters of Eq.~(\ref{eq:gp-mean}) are refitted per coordinate
from its six most recent informative gains, with the decay rate confined to
the interval discussed in Appendix~\ref{app:tail}.  The residual around the
mean is modeled with the kernel of Eq.~(\ref{eq:gp-kernel}).  For training
inputs $X=\{(\bm{n}_i,a_i)\}$ with residuals $u_i=y_i-m_{a_i}(\bm{n}_i)$, the
training covariance is
\begin{equation}
 K_y=K+{\rm diag}(v_i) ,
 \label{eq:app-training-covariance}
\end{equation}
with $K$ the matrix of the kernel between training inputs and $v_i$ the
variance assigned to observation $i$.  Ordinary observations carry a common
log-space variance $\sigma_n^2$, inferred with the other hyperparameters.
This variance describes the unresolved discrepancy of the log-gain field,
the HH energies themselves being deterministic.  On the other hand, the
soft-censored observations of Sec.~\ref{sec:acquisition}, namely rank-neutral
refinements and the first negligible probe of a coordinate, carry a much
larger variance, a log-space standard deviation of $2.5$ against the
$\sigma_n$ of ordinary ones, so they provide weak evidence for a small gain
without suppressing the coordinate.

The hyperparameters, given by
\begin{equation}
 \theta=(\log\ell,\log\sigma_{\rm sh},\log\sigma_{\rm ind},\log\sigma_n) ,
 \label{eq:app-hyperparameters}
\end{equation}
minimize the regularized negative log marginal likelihood~\cite{Rasmussen2006}
\begin{equation}
 \mathcal{L}(\theta)
 = \tfrac12 \bm{u}^{T}K_y^{-1}\bm{u}
 + \tfrac12\log\left[\det K_y\right]
 + \tfrac12\|\theta-\theta_0\|^2 ,
 \label{eq:app-marginal-likelihood}
\end{equation}
up to a constant, where $\theta_0$ holds the default values $\ell=1.5$,
$\sigma_{\rm sh}=0.35$, $\sigma_{\rm ind}=0.75$ and $\sigma_n=0.10$, or the
transferred ones, and the quadratic term prevents large excursions on small
samples.  For a set of candidate inputs, denoted by a star, the usual GP
expressions~\cite{Rasmussen2006} give the predictive mean and covariance
\begin{align}
 \bm{\mu}_\star &= \bm{m}_\star+K_{\star X}K_y^{-1}\bm{u} ,
 \label{eq:app-predictive-mean}\\
 \Sigma_\star^{(0)} &= K_{\star\star}-K_{\star X}K_y^{-1}K_{X\star} ,
 \label{eq:app-predictive-covariance}
\end{align}
where $K_{\star X}$ is the matrix of the kernel between candidate and
training inputs, $K_{X\star}$ its transpose, and $K_{\star\star}$ the matrix
of the kernel between candidates.  To the diagonal terms of
$\Sigma_\star^{(0)}$, i.e.\ to the variance of each candidate, $\sigma_n^2$ is
added.

Because the data are gathered adaptively and the sample is small, the nominal
uncertainty is calibrated on leave-one-out errors~\cite{Rasmussen2006}.  With
$Q=K_y^{-1}$ and $\bm{w}=Q\bm{u}$, the standardized residual at training
point $i$ is $s_i=|w_i|/\sqrt{Q_{ii}}$.  For the target coverage $\gamma$ we
take a finite-sample upper quantile $q_\gamma$ of $\{s_i\}$, compare it with
$z_\gamma=\Phi^{-1}[(1+\gamma)/2]$, and scale the covariance by
$s_{\rm cal}^2$ with
\begin{equation}
 s_{\rm cal} = \max\left(1,\,q_\gamma/z_\gamma\right).
 \label{eq:app-calibration-factor}
\end{equation}

For the extrapolated estimate of Sec.~\ref{sec:stopping}, successive
hypothetical refinements are generated along each coordinate up to
$x_a^{\max}$, holding the others fixed, and all their log gains are predicted
jointly.  This extrapolation is regularized, while the one-step acquisition
is not.  An unconstrained residual can make the predicted sequence rise and the
sum diverge, so the predicted log gains of each coordinate are bounded by its own
fitted decay law raised by the upper quantile of the residuals that
coordinate has actually displayed against that law, the headroom kept
between a minimum of $0.35$ and a maximum of $2.0$ in log-gain units.  Monotonicity is
deliberately not imposed: a selection rule can give zero to one shell while
the next carries strength, and forcing the sequence to decrease would
exclude those cases.  Broad uncertainties are damped in a way that
preserves the correlation matrix.  Transforming the joint lognormal to the gain scale,
including its cross covariances, gives the first of the three estimators of
Sec.~\ref{sec:stopping}, which we call $\Delta_{\rm GP}$, with a correlated
uncertainty $\sigma_{\rm GP}$; both enter Eq.~(\ref{eq:app-combination}) of
Appendix~\ref{app:tail}.

\section{\label{app:tail}The remaining tail estimators}

The second estimator fits a decay law to the informative gains of each
coordinate and sums it up to $x_a^{\max}$.  Two families are admitted: the
exponential law, i.e.\ $\log g_a(n)=\alpha_a-\beta_a n$, and the power law,
i.e.\ $\log g_a(n)=\alpha_a-p_a\log n$, with $\beta_a$ and $p_a$ confined to
$0.04\leq\beta_a\leq3$ per shell and $1.05\leq p_a\leq12$, intervals wide
enough to be uninformative but narrow enough to exclude degenerate fits.  The family is chosen per coordinate by the smaller
root-mean-square residual, and only once at least four informative
observations are available; below that the exponential is used.  When a rate
is clipped to its interval the intercept is re-anchored on the last
informative observation, so that a clipped slope is never paired with an
unclipped intercept.  The hyperangular coordinates select the exponential and
the hyperradial ones the power law, consistently across the four baseline runs.  The tail of a coordinate is
the sum of the fitted law from its current shell to $x_a^{\max}$, evaluated in
closed form, and the estimator is the sum over coordinates.

The third estimator uses no parametric form at all.  It collects the consecutive gain
ratios $g_a(n+1)/g_a(n)$ observed anywhere in the trajectory, keeps those
below one, and takes their upper $0.90$ quantile as a single decay ratio;
the tail of each coordinate is then the geometric sum of that ratio from its
latest measured gain to $x_a^{\max}$.  Ratios at or above one, the
reawakenings, are excluded from the ratio fit but remain represented through
the latest gain of their coordinate.  Because the quantile is taken over the
whole trajectory rather than per coordinate, this estimator remains
defined when a coordinate has too few observations to fit.

The three estimators are combined robustly.  Defining
$\{\Delta^{(k)}\}_{k=1,2,3}$ the three estimates, we obtain
\begin{equation}
 \Delta_{K_{\max}}={\rm median}_k\Delta^{(k)}
 \label{eq:app-median}
\end{equation}
and their median absolute deviation as
\begin{equation}
 D={\rm median}_k|\Delta^{(k)}-\Delta_{K_{\max}}| .
 \label{eq:app-mad}
\end{equation}
Finally,
\begin{equation}
 \sigma_{K_{\max}} = \max\left[
 \sqrt{D^{2}+\min\left(\sigma_{\rm GP},\lambda\Delta_{K_{\max}}\right)^{2}},
 \; f\Delta_{K_{\max}}\right] ,
 \label{eq:app-combination}
\end{equation}
with $\lambda=10$ and a relative floor $f=0.25$, raised to $0.50$ on fewer
than eight training points.  These values are our choice, fixed as described
at the beginning of Appendix~\ref{app:gp}.  The limit $\lambda$ matters when one estimator
degenerates: an uncertainty orders of magnitude larger than the quantity
it qualifies signals that the estimator has become uninformative, and
without the limit it would set the bound.  The floor $f$ reflects that
none of the three is reliable to better than a fraction of itself.  The tail is summed only over coordinates the
model can still reach, since one that never becomes admissible is never
probed and would otherwise contribute its optimistic prior.  The conformal
scale $s_{\rm conf}$ that multiplies the result is defined with
Eq.~(\ref{eq:certified-bound}).

\bibliography{apssamp}

\end{document}